\documentclass[conference]{sig-alternate}
\usepackage{cite}
\usepackage{amsmath,amssymb,amsfonts, physics}
\usepackage{algorithmic}
\usepackage{graphicx}
\usepackage{textcomp}
\usepackage{xcolor}
\usepackage{fancyhdr}
\usepackage[hyphens]{url}
\usepackage{xspace}
\usepackage{graphics}

\usepackage{microtype}
\usepackage{graphicx}
\usepackage{subfigure}
\usepackage{booktabs} 

\usepackage{adjustbox}
\usepackage{array}
\usepackage{multirow}
\usepackage[skip=5pt]{caption}
\usepackage{float}
\usepackage{algorithm2e}
\SetAlFnt{\small}

\newcommand{\lpfixme}[1]{\textcolor{black}{#1}}

\newcommand\blfootnote[1]{%
  \begingroup
  \renewcommand\thefootnote{}\footnote{#1}%
  \addtocounter{footnote}{-1}%
  \endgroup
}

\def\BibTeX{{\rm B\kern-.05em{\sc i\kern-.025em b}\kern-.08em
    T\kern-.1667em\lower.7ex\hbox{E}\kern-.125emX}}

\newcommand{\NAME}{EdgeBERT\xspace}

\newcommand{\iscasubmissionnumber}{914}

\fancypagestyle{firstpage}{
  \fancyhf{}

  \fancyhead[C]{\normalsize{ISCA 2021 Submission
      \textbf{\#\iscasubmissionnumber} \\ Confidential Draft: DO NOT DISTRIBUTE}} 
  \fancyfoot[C]{\thepage}
}

\title{\NAME: Sentence-Level Energy Optimizations for Latency-Aware Multi-Task NLP Inference\vspace{-2.5em}}
\author{Thierry Tambe$^1$, Coleman Hooper$^1$, Lillian Pentecost$^1$, Tianyu Jia$^1$, \\ En-Yu Yang$^1$, Marco Donato$^2$, Victor Sanh$^3$, Paul N. Whatmough$^{4,1}$, \\ Alexander M. Rush$^{5,3}$, David Brooks$^1$, Gu-Yeon Wei$^1$ \\
{\normalsize{} $^1$Harvard University, $^2$Tufts University, $^3$Hugging Face, $^4$Arm Research, $^5$Cornell University\par}}

\begin{document}
\maketitle
\pagestyle{plain}


\begin{abstract}

Transformer-based language models such as BERT provide significant accuracy improvement to a multitude of natural language processing (NLP) tasks. However, their hefty computational and memory demands make them challenging to deploy to resource-constrained edge platforms with strict latency requirements. 

We present \NAME, an in-depth algorithm-hardware co-design for latency-aware energy optimizations for multi-task NLP. 
\NAME employs entropy-based early exit predication in order to perform dynamic voltage-frequency scaling (DVFS), at a sentence granularity, for minimal energy consumption while adhering to a prescribed target latency. Computation and memory footprint overheads are further alleviated by employing a calibrated combination of adaptive attention span, selective network pruning, and floating-point quantization.

Furthermore, in order to maximize the synergistic benefits of these algorithms in always-on and intermediate edge computing settings, we specialize a 12nm scalable hardware accelerator system, integrating a fast-switching low-dropout voltage regulator (LDO), an all-digital phase-locked loop (ADPLL), as well as, high-density embedded non-volatile memories (eNVMs) wherein the sparse floating-point bit encodings of the shared multi-task parameters are carefully stored.
\blfootnote{To appear in 54th IEEE/ACM International Symposium on Microarchitecture (MICRO 2021)}
Altogether, latency-aware multi-task NLP inference acceleration on the \NAME hardware system generates up to 7$\times$, 2.5$\times$, and 53$\times$ lower energy compared to the conventional inference without early stopping, the latency-unbounded early exit approach, and CUDA adaptations on an Nvidia Jetson Tegra X2 mobile GPU, respectively.

\end{abstract}

\section{Introduction}\label{intro}
Transformer-based networks trained with large multi-domain datasets have unlocked a series of breakthroughs in natural language learning and representation. A major catalyst of this success is the \textit{Bidirectional Encoder Representations from Transformers} technique, or BERT~\cite{bert_nlp}, which substantially advanced nuance and context understanding. 
Its pre-training strategy, which consists of learning intentionally hidden sections of text, have proven beneficial for several downstream natural language processing (NLP) tasks.
BERT has sparked leading-edge performance in NLP leaderboards~\cite{gluebench, squad}, and it is now applied at a global scale in web search engines~\cite{bert_google_search} with marked improvements in the quality of query results.

\begin{figure}[t!]
    \centering
    \includegraphics[width=1.01\linewidth]{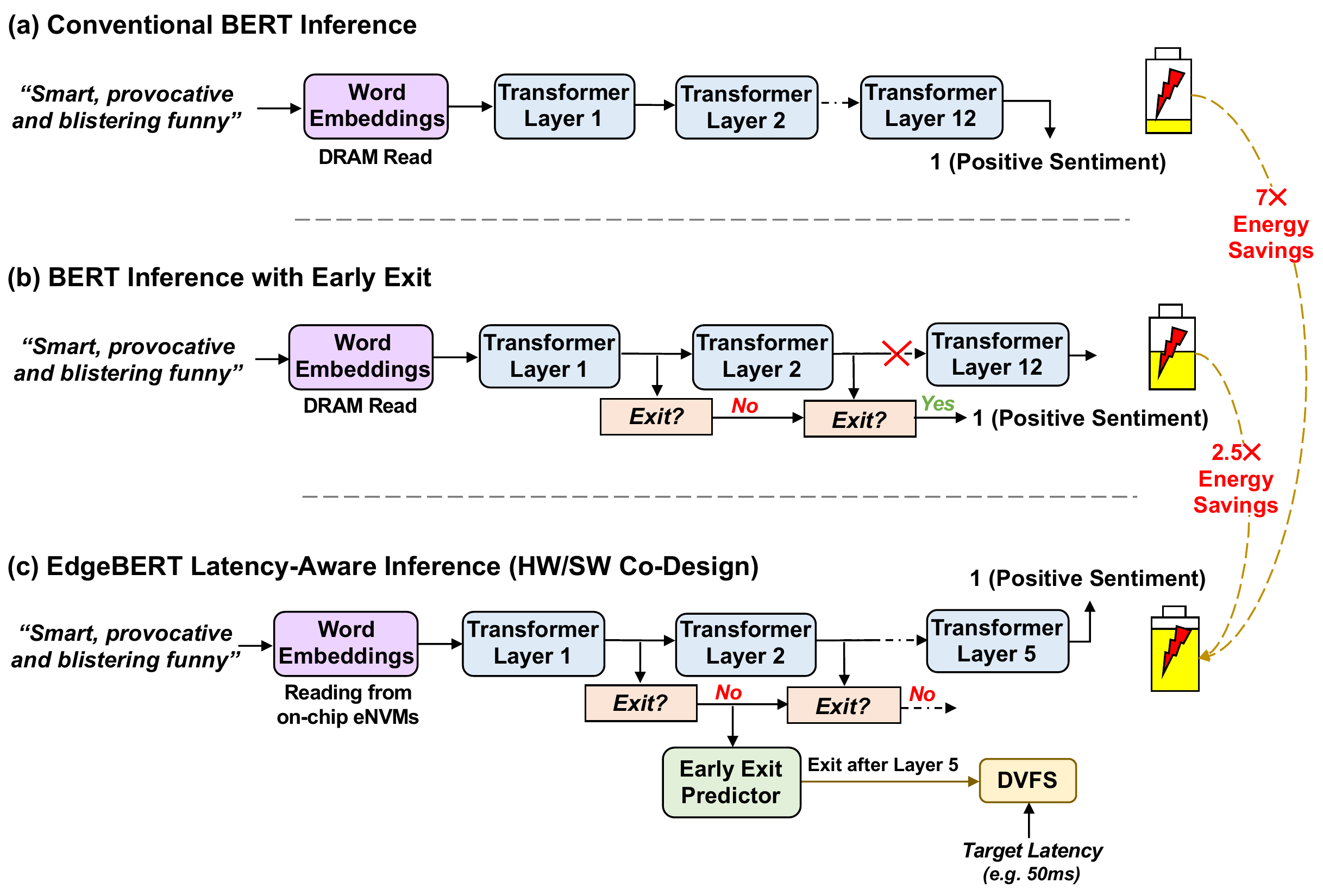}
\caption{(a) Conventional BERT inference, (b) Conventional latency-unbounded BERT inference with early exit. (c) Proposed latency-bounded inference. The entropy result from the first layer is used to auto-adjust the accelerator supply voltage and clock frequency for energy-optimal operation while meeting an application end-to-end latency target.}
\label{fig:ee_proposal}
\end{figure}

Advances in NLP models are also fueling the growth of intelligent virtual assistants, which leverage NLP to implement interactive voice interfaces. 
Currently, these applications are offloaded from the edge device to the cloud.
However, they are naturally better suited to deployment on edge devices, where personal data can be kept private and the round trip latency to the cloud is removed.
However, the impressive performance of BERT comes with a heavy compute and memory cost, which makes on-device inference prohibitive.
Most significantly, the BERT base model consumes a staggering 432 MB of memory in native 32-bit floating-point (FP32).

Therefore, the goal of deploying BERT on edge/mobile devices is challenging and requires tight co-design of the BERT model optimizations with dedicated hardware acceleration and memory system design.
The constraints on mobile can be quite different to the datacenter scenario, where BERT has been mainly deployed to date.
Firstly, since we are dealing with user input, we need to meet real time throughput requirements to prevent a noticeable lag to the user.
Secondly, energy consumption is a critical concern on mobile devices, both for the model inference and also the associated data movement cost.
A number of prior works have been proposed to reduce BERT storage and computation overheads~\cite{compress_transformers}. In fact, most of the compression techniques (weight pruning~\cite{bert_prune}, distillation~\cite{distilbert}, quantization~\cite{bert_q8bit,qbert}) originally proposed for convolutional and recurrent neural networks (CNNs, RNNs) have been independently applied to Transformer-based DNNs.

In this work, we present {\it EdgeBERT}, a principled latency-driven approach to accelerate NLP workloads with minimal energy consumption via early exit prediction, dynamic voltage-frequency scaling (DFVS), and non-volatile memory bitmask encoding of the shared word embeddings. 
In conventional BERT inference (Fig.~\ref{fig:ee_proposal}{a}), the final classification result is generated by the last Transformer layer.
Early exit mechanisms~\cite{DeeBERT,patience_bert,branchynet,the_right_tool} (Fig.~\ref{fig:ee_proposal}(b)) have been proposed to reduce the average energy and latency.
The early exit entropy, which is a probabilistic measure of the classification confidence, is evaluated at the output of each computed Transformer layer and the inference exits when the entropy value falls below a pre-defined threshold. 
While this approach can appreciably reduce computation and energy costs, the achieved latency can vary drastically from one input sentence to another, potentially violating the strict real time latency constraint of the application.
In contrast, \NAME uses this upper-bound latency and the target entropy as optimization constraints, and then dynamically auto-adjusts the accelerator supply voltage and clock frequency to minimize energy consumption (Fig.~\ref{fig:ee_proposal}(c)), while meeting the real time throughput requirement. 
Since energy scales quadratically with $V_{DD}$ and linearly with the number of computation cycles, our DVFS algorithm finds the lowest possible frequency/voltage, while also minimizing the total number of FLOPs via adaptive attention span predication. 

While the benefits of early exit and attention predications can be reaped on commodity GPUs, we unlock additional energy savings by co-designing the hardware datapaths. Specifically, we exploit these algorithmic optimizations in the \NAME accelerator system, which integrates a fast-switching low-dropout (LDO) voltage regulator and an all-digital phase-locked loop (ADPLL) for DVFS adjustments. The \NAME accelerator uses bit-mask encoding for compressed sparse computations, while optimizing key operations (entropy assessment, layer normalization, softmax and attention masking) for numerical stability and energy efficiency.

Furthermore, edge/IoT devices operate intermittently which motivates powering down as much as possible.  The model's weights, typically stored in on-chip SRAMs, either have to be reloaded from DRAM each wake up cycle or the on-chip SRAMs storing the weights must be kept on, wasting leakage power~\cite{dnn-mem}. Embedded non-volatile memories (eNVMs), which have shown considerable progress in recent years,  offer great promise, if used judiciously, to eliminate the power penalty associated with intermittent operation. For this purpose, we perform monte-carlo fault injection simulations to identify robust and viable eNVM structures for storing the shared NLP multi-task parameters with bitmask encoding. Our resulting eNVM configuration significantly alleviates the energy and latency costs associated with multi-task intermediate computing by as much as 66,000$\times$ and 50$\times$, respectively.

\begin{figure}[!t]
    \centering
    \includegraphics[width=\columnwidth]{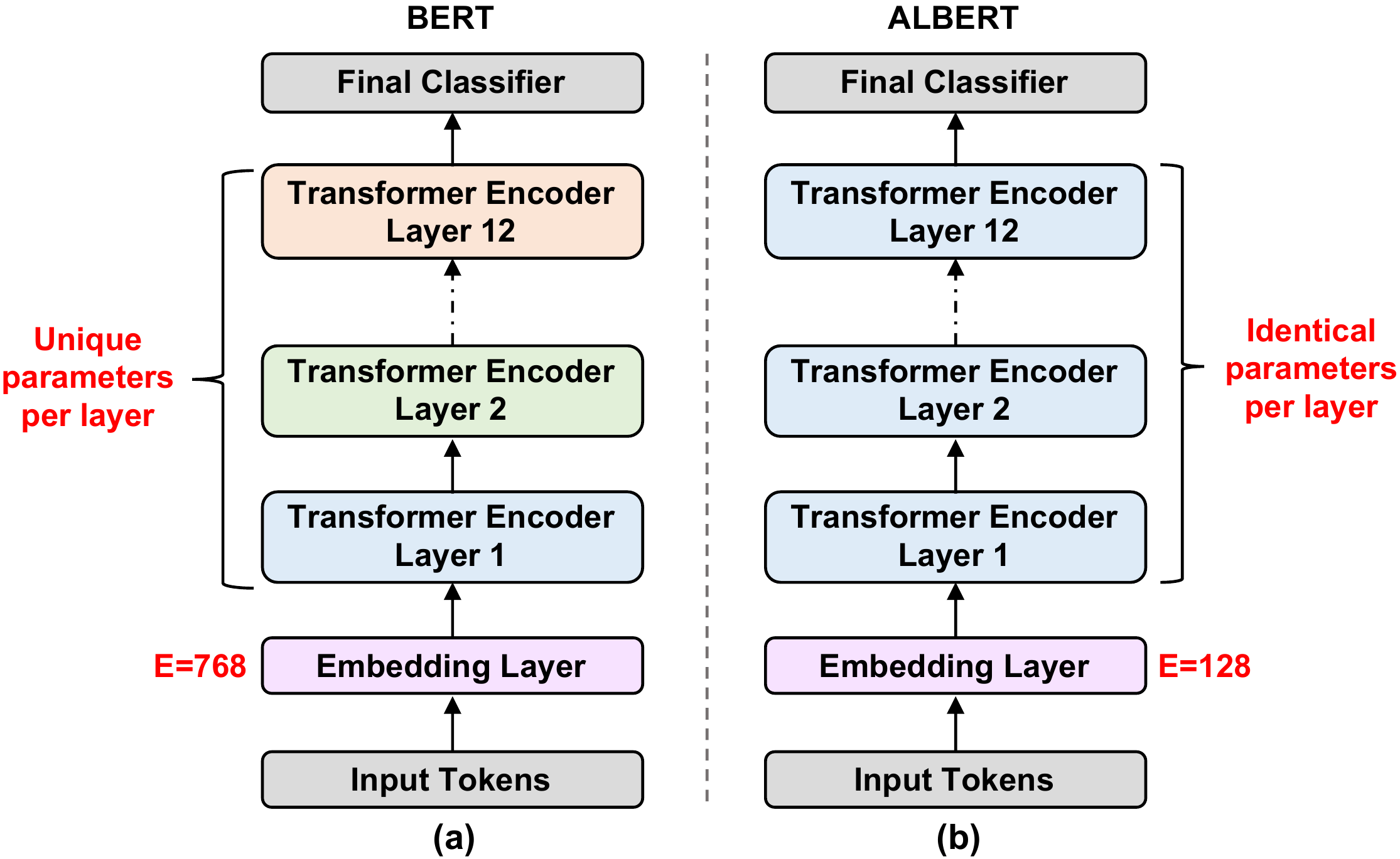}
    \caption{Comparison between (a) BERT, and (b) ALBERT base models. ALBERT uses a smaller embedding size and its Transformer encoder layers share the same parameters.}
    \label{fig:bert_albert_compare}
\end{figure}

Altogether, \NAME generates on average up to 7$\times$, and 2.5$\times$ per-sentence energy savings compared to the conventional BERT inference, and latency-unaware early exit approaches, respectively.

This paper therefore makes the following contributions:
\begin{itemize} 
\item We propose \NAME, a novel algorithm-hardware co-design approach to enable 
latency-bound NLP workloads on resource-constrained embedded devices. 



\item Recognizing that BERT word embeddings are shared across NLP tasks, we significantly alleviate off-chip communication costs by identifying viable and robust multi-level eNVM structures for storing the multi-task word embeddings. 

\item Leveraging the insights from this broad analysis, we propose and design a 12nm accelerator that integrates a fast-switching LDO, an ADPLL, and a compressed sparse hardware accelerator that efficiently computes the DVFS, entropy, and adaptive attention span predication algorithms and other key Transformer operations using specialized datapaths. 

\item We evaluate the energy consumption of latency-bound inference on four NLP tasks, and find that the \NAME hardware accelerator system generates up to 7$\times$, 2.5$\times$, and 53$\times$ lower energy compared to the unoptimized baseline inference without early exit, the conventional latency-unaware early exit approach, and CUDA adaptations on an Nvidia Jetson Tegra X2 mobile GPU respectively.  

\end{itemize}

\section{Background }\label{sec:background}

\begin{figure*}[t]
    \centering
    \includegraphics[width=0.98\linewidth]{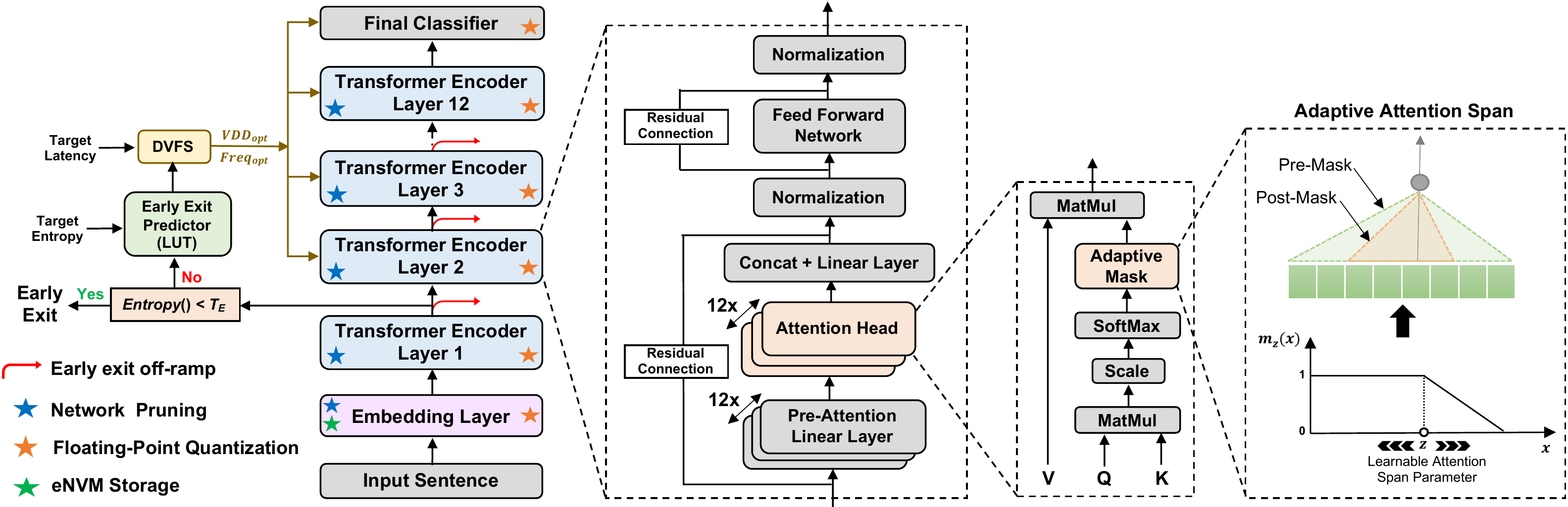}
    \caption{Memory and latency optimizations incorporated in the \NAME methodology.  Each self-attention head learns its own optimal attention span. Network pruning is performed on all Transformer encoders. The embedding layer is stored in non-volatile memory. Floating-point quantization is applied to all weights and activations. During real-time on-device execution, DVFS is performed for latency-bounded inference.}
    \label{fig:albert_opts}
\end{figure*}

\subsection{Benchmarks}
The General Language Understanding Evaluation (GLUE) benchmark is the most widely used tool to evaluate NLP performance. It consists of nine English sentence understanding tasks covering three categories: Single-Sentence, Similarity and Paraphrase, and Inference~\cite{gluebench}. This collection of datasets is specifically designed to favor models that can adapt to a variety of NLP tasks. To validate the robustness and generalization performance of the \NAME methodology, we conduct our evaluation on the four GLUE tasks with the largest corpora, which cover all three GLUE categories: SST-2 (Single-Sentence), QQP (Similarity and Paraphrase), and QNLI and MNLI (Inference).

\subsection{Variations of BERT}
Since the advent of BERT with 110M parameters, a number of variants were proposed to alleviate its memory consumption or to further improve its prediction metrics. RoBERTa~\cite{roberta} generalizes better on several GLUE tasks by training on significantly more data, and for a longer amount of time, but remains as computationally intensive as BERT. DistilBERT~\cite{distilbert} and MobileBERT~\cite{mobilebert} leverage knowledge distillation to reduce BERT size by 1.7$\times$ and 4.3$\times$, respectively, with iso-accuracy. SqueezeBERT~\cite{squeezebert} substitutes several operations in the Transformer encoder with 1D grouped convolutions achieving 4$\times$ speedup while being 2$\times$ smaller. Q8BERT~\cite{bert_q8bit} employs a symmetric linear quantization scheme for quantizing both weights and activations into 8-bit integers. In contrast, in this work we leverage the higher dynamic range of floating-point encodings for greater quantization resilience.
ALBERT~\cite{albert} yields the smallest footprint to date for a compressed BERT variant with only 12M parameters, with competitive accuracy on the GLUE benchmarks.

Fig.~\ref{fig:bert_albert_compare} summarizes the key differences between the ALBERT model and the base BERT model. While each of BERT's twelve encoder layers have a unique set of weights, ALBERT's encoder layers instead share and reuse the same parameters -- resulting in significant compression. The encoder block in both models has the same architecture as the legacy Transformer network~\cite{attention_is_all_you_need}, but with twelve parallel self-attention heads. Moreover, ALBERT employs a smaller embedding size (128 vs. 768) thanks to factorization in the embedding layer. In this work, we adopt the ALBERT variant as an efficient baseline. This work further pursues strategies to reduce latency and storage requirements to suit embedded hardware platform constraints.

\section{Alleviating Transformer memory and computation costs} \label{sec:meth}

An accelerator's energy consumption can be abstracted as:
\[Energy\propto \alpha C V_{DD}^2  N_{cycles}\] where $\alpha$, $C$, $V_{DD}$ and $N_{cycles}$ are the switching activity factor, the effective wire and device capacitance, the supply voltage, and the required number of clock cycles to complete the inference, respectively. While the DVFS algorithm (Sec.~\ref{sec:dvfs_sw}) lowers the energy quadratically by bringing $V_{DD}$ down to the lowest optimal voltage, in this section, we explore avenues to further reduce the energy by minimizing $\alpha$, $C$, and $N_{cycles}$. 

For this purpose, we carefully incorporate into the multi-task ALBERT inference: 1) adaptive attention span predication and early exit which reduce $N_{cycles}$; 2) network pruning, which ultimately reduces $\alpha$; and 3) floating-point quantization helping decrease $C$, altogether with minimal accuracy degradation. While briefly describing these optimizations individually in this section, we provide a reasoned methodology for applying them to the ALBERT model, as shown in Fig.~\ref{fig:albert_opts}.

\subsection{Entropy-based Early Exit}
The motivation behind early exit (EE) is to match linguistically complex sentences with larger (or deeper) models and simple sentences with smaller (or shallower) models~\cite{DeeBERT, dac19}. This is typically done by adding a lightweight classifier at the output of the Transformer layer so that a given input can exit inference earlier or later in the stack, depending on its structural and contextual complexity. The classifier computes and compares the entropy of an output distribution with a preset ``confidence" threshold, $E_{T}$, in order to assess whether the prediction should exit or continue inference in the next Transformer encoder layer. 
The entropy metric quantifies the amount of uncertainty in the data. Smaller entropy values at a Transformer layer output implies greater confidence in the correctness of the classification result. 
The entropy \textit{H} on sample \textit{x} is estimated as: 
\begin{equation}\label{eq:entropy}
\resizebox{0.80\hsize}{!}{%
$H(x) = - \sum p(x)\log p(x) 
= \ln (\sum\limits_{k=1}^{n} e^{x_k}) -
\frac{
  \sum\limits_{k=1}^{n} x_k e^{x_k}
}{\sum\limits_{k=1}^{n} e^{x_k}}$%
}
\end{equation} 
The early exit condition is met when $H(x)$ $<$ $E_{T}$. Therefore, the larger $E_{T}$ becomes, the earlier the sample will exit (i.e. $N_{cycles}$ becomes smaller) with potentially lower accuracy.

In this work, we modify the conventional EE inference approach by predicting the early exit layer from the output of the first Transformer layer in order to run the rest of the network computation in an energy-optimal and latency-bounded manner (Sec.~\ref{sec:ep_dvfs}).

\subsection{Adaptive Attention Span}

The attention mechanism~\cite{bahdanau_attention} is a powerful technique that allows neural networks to emphasize the most relevant tokens of information when making predictions. The base ALBERT model contains up to twelve parallel attention heads -- each learning their own saliency weights on the full length of the encoder input. However, depending on the complexity of the task, many heads can be redundant and can be safely removed without impacting accuracy~\cite{sixteen_heads}. Furthermore, the cost of computing the attention mechanism scales quadratically with the sequence length. Therefore, there is potentially a meaningful amount of computations and energy to be saved in optimizing the inspection reach of every attention head.

In the quest to avoid needless attention computations in ALBERT, a learnable parameter \textit{z} is introduced in the datapath of each self-attention head in order to find its own optimal attention span~\cite{adaptive_attn_span}. The parameter \textit{z} is mapped to a masking function with a [0, 1] output range, as shown in Fig.~\ref{fig:albert_opts}. The masked span is then applied on the attention weights in order to re-modulate their saliencies. The optimal span is automatically learned during the fine-tuning process by adding back the average loss from the reduced span to the training cross-entropy loss.

The maximum sentence length for fine-tuning the GLUE tasks is 128. As a result, shorter sentences are typically zero-padded to 128 during the tokenization pre-processing.  Table~\ref{tab:attn_head} shows the final attention span learned by each self-attention head when fine-tuning with the adaptive attention span technique. Strikingly, the twelve parallel self-attention heads in ALBERT do not need to inspect their inputs at maximum span. In fact, more than half of the attention heads, 8 for MNLI and QQP and 7 for SST-2 and QNLI, can be completely turned off with minimal accuracy loss.  This amounts to a 1.22$\times$ and 1.18$\times$ reduction, respectively, in the total number of FLOPS (which linearly correlates with $N_{cycles}$) required for single-batch inference.

\begin{table}[!t]
\caption{Learned spans of every attention head in ALBERT. \\ {Baseline Acc: MNLI=85.16, QQP=90.76, SST-2=92.20, QNLI=89.48} }
 \label{tab:attn_head}
   \centering
 \resizebox{\columnwidth}{!}{
\begin{tabular}{c|c|c|c|c|c|c|c|c|c|c|c|c|c|c|c|}
\cline{2-16}
                            & \multicolumn{12}{c|}{Attention Head \#}               & \multirow{2}{*}{\begin{tabular}[c]{@{}c@{}}Avg. \\ Span\end{tabular}} & \multirow{2}{*}{Acc.} & \multirow{2}{*}{Diff.} \\ \cline{2-13}
                            & 1  & 2 & 3 & 4 & 5 & 6   & 7  & 8  & 9 & 10 & 11 & 12 &                                                                       &                       &                        \\ \hline
\multicolumn{1}{|c|}{MNLI}  & 20 & 0 & 0 & 0 & 0 & 0   & 36 & 81 & 0 & 0  & 0  & 10 & 12.3                                                                  & 85.11                 & -0.05                  \\ \hline
\multicolumn{1}{|c|}{QQP}   & 16 & 0 & 0 & 0 & 0 & 0   & 40 & 75 & 0 & 0  & 0  & 2  & 11.0                                                                  & 90.80                 & 0.04                   \\ \hline
\multicolumn{1}{|c|}{SST-2} & 31 & 0 & 0 & 0 & 0 & 101 & 14 & 5  & 0 & 36 & 0  & 0  & 15.6                                                                  & 91.99                 & -0.21                  \\ \hline
\multicolumn{1}{|c|}{QNLI}  & 39 & 0 & 0 & 0 & 0 & 105 & 22 & 19 & 0 & 51 & 0  & 0  & 19.6                                                                  & 88.92                 & -0.56                  \\ \hline
\end{tabular}
}
\end{table}

The twelve attention spans, learned during fine-tuning, are written to registers in the \NAME accelerator in the form of a 128-wide vector -- in order to predicate on the inference computation of the multi-head attention. Notably, all the computations inside any of the twelve attention head units can be effectively skipped in case its associated attention span mask is 100\% null. The \NAME accelerator takes advantage of this observation in a proactive manner during inference in the custom hardware (Sec. \ref{sec:mha}). 

\subsection{Network Pruning}
The \NAME hardware accelerator (Sec.~\ref{sec:arch}) executes sparse computations and saves energy by gating MACs whenever input operands are null. Therefore, the extent to which we can prune the ALBERT model, without appreciable accuracy loss, determines the overall accelerator energy efficiency. 

In this work, we consider both movement pruning~\cite{mvp} and the well-known magnitude pruning~\cite{deepcompression} methods. 
Movement pruning is a first-order pruning technique that is applied during model fine-tuning which eliminates weights that are dynamically shrinking towards 0 (i.e., according to the movement of the values). In some cases, magnitude pruning may be a sub-optimal method to use during transfer learning, as pre-trained weights closer to zero may have a high chance of being eliminated regardless of the fine-tuning requirement. 
We observe that movement pruning particularly outperforms magnitude-based pruning in high sparsity regimes, as each individual remaining weight becomes more important to learn the task at hand.
Therefore, choosing between the two pruning techniques would depend on the per-task tolerance to increasing sparsity levels. 
We note that magnitude pruning is always applied to the ALBERT embedding layer in order to enforce uniformity in the data during multi-domain on-chip acceleration -- as using movement pruning on the embedding layer would make its weights unique for each NLP domain, thereby forgoing opportunities for data reuse in hardware.

\subsection{Floating-Point Quantization}
DNN algorithmic resilience allows for parameters to be represented in lower bit precision without accuracy loss. Fixed-point or integer quantization techniques, commonly adopted in CNN models, suffer from limited range and may be inadequate for NLP models, whose weights can be more than an order of magnitude larger~\cite{adaptivfloat}. 
This phenomenon is owed to layer normalization~\cite{layernorm}, which is commonly adopted in NLP models and has invariance properties that do not reparameterize the network -- unlike batch normalization~\cite{batchnorm}, which produces a weight normalization side effect in CNNs. 

In this work, we employ floating-point based quantization, which provides 2$\times$ higher dynamic range compared to integer datatypes~\cite{rethinkfloat}. Both weights and activations are quantized across ALBERT layers to 8-bit precision. We also performed a search on the optimal exponent bit width to satisfy the dynamic range requirements of the ALBERT model. Setting the floating-point exponent space to 4 bits within the 8-bit word size, with the exponent being scaled at a per-layer granularity, provided the best accuracy performance across NLP tasks.

\section{Non-Volatile Memory Storage of Shared Parameters} \label{sec:nvm}


In contrast to task-specific encoder weights, word embedding parameters are deliberately fixed during fine-tuning and reused across different NLP tasks. We seek to avoid the energy and latency costs of reloading the word embeddings from off-chip memory for different tasks by storing these shared parameters in embedded non-volatile memories (eNVMs). eNVM storage also enables energy-efficient intermittent computing because the embedding weights will be retained if and when the system-on-chip powers off between inferences.
However, despite their compelling storage density and read characteristics, eNVMs exhibit two main drawbacks: potentially high write cost (in terms of energy and latency) and decreased reliability, particularly in multi-level cell (MLC) configurations~\cite{rram_dac}. 
Fortunately, the word embeddings are acting as read-only parameters on-chip, which makes them highly suitable for eNVM storage, but previous work highlights the need to study the impacts of faulty, highly-dense ReRAM storage on DNN task accuracy~\cite{maxnvm}. On the other hand, encoder weights need to be updated when switching across different NLP tasks. To prevent the energy and latency degradation that would follow from updating the encoder weight values in eNVMs, we map the natural partition of shared and task-specific parameters to eNVMs and SRAMs, respectively~\cite{memti}.
\subsection{eNVM Modeling Methodology}
\begin{table}[!t]
\caption{Results of fault injection simulations modeling impact of ReRAM embedding storage on task accuracy. SLC=single-level cell (1 bit per cell). MLC2= 2 bits per cell. MLC3 = 3 bits per cell. }
\label{tab:nvm_table}
\begin{center}
 \begin{small}
 \begin{sc}
 
 \resizebox{\columnwidth}{!}{
\begin{tabular}{c|c|c|c|c|c|c|}
\cline{2-7}
                                                                                      & \multicolumn{2}{c|}{SLC}  & \multicolumn{2}{c|}{MLC2} & \multicolumn{2}{c|}{MLC3} \\ \cline{2-7} 
                                                                                      & mean         & min        & mean        & min         & mean        & min         \\ \hline
\multicolumn{1}{|c|}{MNLI}                                                            & 85.44        & 85.44      & 85.44       & 85.44       & 85.42       & 85.25       \\ \hline
\multicolumn{1}{|c|}{QQP}                                                             & 90.77        & 90.77      & 90.77       & 90.77       & 90.75       & 90.61       \\ \hline
\multicolumn{1}{|c|}{SST-2}                                                           & 92.32        & 92.32      & 92.32       & 92.32       & 91.86       & 90.83       \\ \hline
\multicolumn{1}{|c|}{QNLI}                                                            & 89.53        & 89.53      & 89.53       & 89.53       & $\textbf{88.32}$       & $\textbf{53.43}$       \\ \hline
\multicolumn{1}{|c|}{\begin{tabular}[c]{@{}c@{}}Area Density\\ ($mm^2$/MB)\end{tabular}} & \multicolumn{2}{c|}{0.28} & \multicolumn{2}{c|}{0.08} & \multicolumn{2}{c|}{0.04} \\ \hline
\multicolumn{1}{|c|}{\begin{tabular}[c]{@{}c@{}}Read Latency\\ ($ns$)\end{tabular}}     & \multicolumn{2}{c|}{1.21} & \multicolumn{2}{c|}{1.54} & \multicolumn{2}{c|}{2.96} \\ \hline
\end{tabular}
}
\end{sc}
\end{small}
\end{center}
\end{table}
This work specifically considers dense, energy-efficient Resistive RAM (ReRAM) arrays~\cite{rram_isscc,rram_isscc2} as an on-chip storage solution for shared embedding parameters. We selected ReRAMs for their relative maturity and demonstrated read characteristics. However, we note that there is a larger design space of opportunities to be explored with other emerging MLC-capable NVM technologies such as PCM~\cite{pcm_nvm}, but is beyond the scope of this work.

We evaluate the robustness of storing the 8-bit quantized word embeddings in eNVM storage. 
In order to quantify the trade-offs between storage density and task accuracy, we use cell characteristics of 28nm ReRAM programmed with varying number of bits per cell~\cite{rram_dac}, and evaluate 100 fault injection trials per storage configuration to identify robust eNVM storage solutions. We leverage and extend Ares~\cite{ares}, which is an existing open-source fault injection framework for quantifying the resilience of DNNs. 

After pruning, we store non-zero compressed embedding weights using a bitmask-style sparse encoding. Previous work demonstrates that DNN weight bitmask values are vulnerable to MLC faults, so the bitmask is protectively stored in lower-risk SLC devices, while we experiment with MLC storage for the non-zero data values \cite{maxnvm}.
\subsection{Optimal eNVM Configuration}
Table~\ref{tab:nvm_table} uncovers exceptional resilience to storing word embeddings in MLC ReRAM. Across many fault injection trials, we observe that MLC2 (ReRAM programmed at 2 bits-per-cell) does not degrade accuracy across multiple tasks, while MLC3 exhibits potentially catastrophic degradation in minimum accuracy and an appreciable decline in average accuracy for the QNLI task, highlighted in bold.
Based on this observation, the \NAME accelerator system leverages MLC2 ReRAMs for word embedding storage (Sec.\ref{sec:arch}).
\setlength{\textfloatsep}{0pt}

\begin{algorithm}[!t]

\small
\scriptsize
\algsetup{linenosize=\tiny}
\DontPrintSemicolon
\KwIn{$E_T$ := target entropy}





\For{input sentence $i = 0$ to $n$}{
    \For{encoder layer l = 1 to 12}{
        $z_l = f(x;\theta | VDD_{nom}, Freq_{max})$\;  
           \If {$entropy(z_l) < E_T$} {
            \textbf{exit inference}\;
        }      
    }
    
}
\caption{{Conventional early exit inference}}
\label{algo:ee}

\end{algorithm}

\section{EdgeBERT's Latency-Aware Inference} \label{sec:ep_dvfs}

The conventional BERT inference (Algorithm~\ref{algo:ee}) with early exit (EE) can significantly reduce BERT inference latency.
To further reduce the energy consumption for NLP inference, a latency-aware inference scheme leveraging the EE predictor and dynamic voltage and frequency scaling (DVFS) is proposed to minimize end-to-end per-sentence energy consumption while satisfying the real-time latency target.

\setlength{\textfloatsep}{0pt}
\setlength{\intextsep}{1\baselineskip}
\begin{algorithm}[!t]

\small
\scriptsize
\algsetup{linenosize=\tiny}
\DontPrintSemicolon
\KwIn{$T$ := per-sentence latency target, $E_T$ := entropy target}





$N_{cycles} := \text{number of clock cycles to compute the Transformer encoder}$

\vspace{5pt}


\For{input sentence $i = 1$ to $n$}{


    
    \For{encoder layer l = 1}{
        $z_l = f(x;\theta | VDD_{nom}, Freq_{max})$\;
        \If {$entropy(z_l) < E_T$} {
            \textbf{exit inference}\;
        } 
        \Else {
            $L_{predict} = LUT(entropy(z_1), E_T)$\;
            $VDD_{opt}, Freq_{opt} = DVFS(L_{predict},T)$\;
        }
    }

    \For{encoder layer $l = 2$ to $L_{predict}$}{
        $z_l = f(x;\theta | VDD_{opt}, Freq_{opt})$\;    
           \If {$entropy(z_l) < E_T$} {
            \textbf{exit inference}\;
        }      
    }
    \textbf{exit inference}\;
    
}
\caption{{\NAME latency-aware inference. Computations exit at the predicted exit layer or earlier.}}
\label{algo:edgebert_hard}
\end{algorithm}
\subsection{Methodology}

DVFS is a widely used technique to dynamically scale down the voltage and frequency for less computationally intensive workloads. In the past, DVFS has been widely deployed in commercial CPUs \cite{dvfs_isscc}, \cite{2021isscc_mediatek_dvfs} and GPUs \cite{2018isscc_intel_dvfs}. However, these schemes typically adjust the voltage and frequency at a coarse granularity at workload-level. In the era of AI, DVFS has started to be explored for DNN accelerators~\cite{sm3_jssc19}. For example, a recent state-of-the-art AI chip has reported per-layer DVFS to save energy \cite{2021isscc_ibm_ml}. In this work, we explore a fine-grained sentence-level DVFS to reduce the energy consumption for NLP inference while meeting the latency target.

The proposed early exit -based latency-aware inference methodology is illustrated in Algorithm~\ref{algo:edgebert_hard}.
The inference of a sentence starts at nominal voltage and maximum frequency, and the entropy value is calculated at the output of the first Transformer encoder layer. The entropy result is then sent to a trained classifier (EE predictor) to predict which following encoder layer should early exit (e.g. early exit at encoder layer 6 before the final encoder layer 12). Based on the predicted early exit layer, the voltage and frequency is scaled down to proper energy-optimal setting for the rest of encoder layers (e.g. encoder layer 2 to 6) while meeting the latency target for each sentence. This scheme produces a quadratic reduction in the accelerator power consumption.

In our work, the EE predictor is a ReLU-activated {five-layer perceptron} neural network with 64 cells in each of the hidden layers. It takes the entropy of encoder layer 1 as input and forecasts the early exit Transformer layer which has an entropy below the desired threshold. 
The neural network architecture of the EE predictor was empirically searched with the goal of minimizing the difference between the predicted and the true entropy-based exit layer. 
For this purpose, we constructed parallel training and test datasets containing the entropy values at the output of the 12 Transformer layers during evaluation on the GLUE benchmarks.

The EE predictor is distilled as a lookup table (LUT) leading to negligible one-time (per-sentence) computational overhead. Furthermore, implementing the EE predictor as a LUT simplifies its hardware operation. 
As the neural network based LUT is error-prone, it may predict a higher exit layer than necessary. Therefore, during the inference, the entropy is checked after each encoder layer for early stopping until the predicted layer. If the computed entropy becomes lower than the exit threshold before the predicted encoder layer, the inference will terminate at that early exit condition point. 
In case the inference reaches the predicted layer, termination occurs even if the entropy at that layer is still higher than the exit threshold in order to not violate timing constraints.

When assessing the impacts of using entropy prediction instead of traditional EE methods, we set a fixed accuracy degradation threshold of 1\%, 2\%, or 5\% (relative to the inference accuracy of the full ALBERT model) and increased the entropy threshold until the accuracy dropped to the desired threshold. This allowed us to compare energy savings between entropy prediction and conventional EE for a fixed accuracy target. For the same accuracy threshold, the entropy threshold for entropy prediction was lower than the entropy threshold for conventional EE, leading to a slightly later average exit layer during inference. However, entropy prediction allows for DVFS since the maximum exit layer is known after the first layer, whereas with the conventional EE appproach, the maximum exit layer is always the final encoder layer. EdgeBERT latency-aware inference therefore achieves greater energy savings than the conventional EE approach by facilitating DVFS (Sec.~\ref{sec:lat_hw_results}).
\begin{figure}[!t]
    \centering
    \includegraphics[width=\columnwidth]{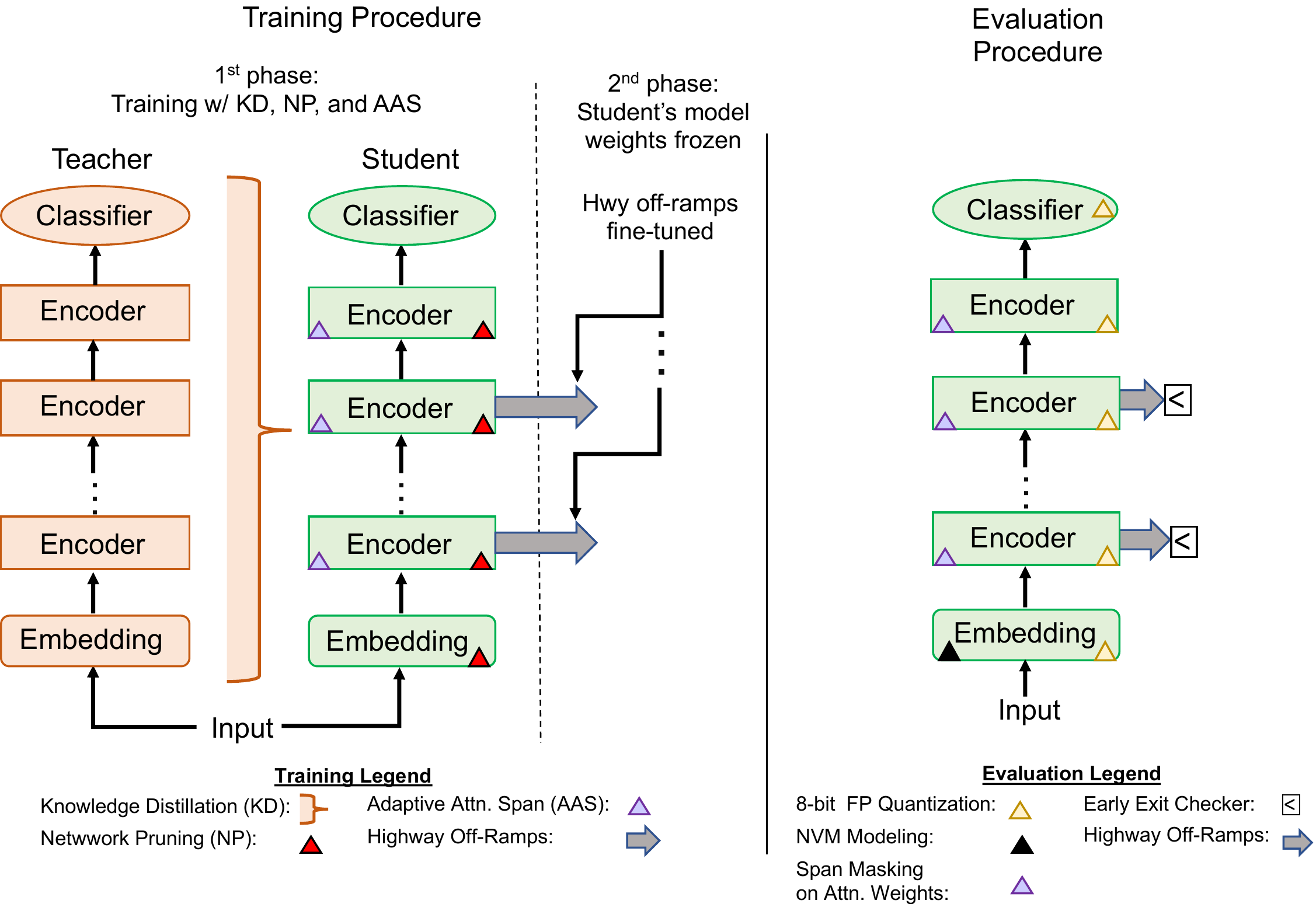}
    \caption{\NAME training and evaluation procedure. 
    }
    \label{fig:train_eval_proc}
\end{figure}
\subsection{On-chip DVFS system} \label{sec:dvfs_sw}
To realize fast per-sentence DVFS, the on-chip DVFS system is developed and integrated within EdgeBERT. The DVFS system includes a DVFS controller, an on-chip synthesizable linear voltage regulator (LDO), and an all-digital PLL (ADPLL). Compared with the conventional workload-level DVFS \cite{dvfs_isscc}, the proposed scheme adjusts voltage and frequency at a finer-grained sentence-level granularity. Based on the predicted early exit layer from the EE predictor, the required run cycles, $N_{cycles}$, for the rest of the encoder layers before early exit can be known. And, knowing the frontend elapsed time $T_{elapsed}$ up to the EE predictor within the per-sentence latency target $T$, the optimal running frequency can be calculated as follows:
\[Freq_{opt} = N_{cycles} / (T - T_{elapsed}) \] 
Meanwhile, the corresponding energy-optimal supply voltage, $VDD_{opt}$, is selected by the DVFS controller to achieve the lowest operational voltage value at $Freq_{opt}$. In the \NAME accelerator system, this is done via indexing the look-up table containing the ADPLL frequency/voltage sweep coordinates. The DVFS is performed for each real-time sentence inference due to its fast response time; the implementation details are shown in Sec.~\ref{sec:dvfs_hw}. 



\section{Algorithmic Synergy} \label{sec:synergy}
\begin{table}[]
\caption{Summary of optimization results in terms of achievable sparsity, attention span with early exit performance and accuracy implications. 
Baseline Acc: MNLI=85.16, QQP=90.76, SST-2=92.20, QNLI=89.48}
 \label{tab:simult_sw}
  \centering
 \resizebox{\columnwidth}{!}{
\begin{tabular}{ccccc|c|c|c|c|c|}
\cline{6-10}
 &  &  &  &  & \multicolumn{2}{c|}{\begin{tabular}[c]{@{}c@{}}Conventional \\ EE Approach\end{tabular}} & \multicolumn{3}{c|}{\begin{tabular}[c]{@{}c@{}}EdgeBERT Latency-Aware\\ Inference\end{tabular}} \\ \cline{2-10} 
\multicolumn{1}{c|}{} & \multicolumn{1}{c|}{\begin{tabular}[c]{@{}c@{}}Embedding\\ Sparsity \\ (\%)\end{tabular}} & \multicolumn{1}{c|}{\begin{tabular}[c]{@{}c@{}}Encoder \\ Sparsity\\  (\%)\end{tabular}} & \multicolumn{1}{c|}{\begin{tabular}[c]{@{}c@{}}Avg. \\ Attn. \\ Span\end{tabular}} & \begin{tabular}[c]{@{}c@{}}Pct. Pt. \\ Acc. Drop\end{tabular} & \begin{tabular}[c]{@{}c@{}}Entropy\\ Threshold\end{tabular} & \begin{tabular}[c]{@{}c@{}}Avg. \\ Exit \\ Layer\end{tabular} & \begin{tabular}[c]{@{}c@{}}Entropy\\ Threshold\end{tabular} & \begin{tabular}[c]{@{}c@{}}Avg.\\ Predicted \\ Exit Layer\end{tabular} & \begin{tabular}[c]{@{}c@{}}Avg. \\ Actual \\ Exit Layer\end{tabular} \\ \hline
\multicolumn{1}{|c|}{\multirow{3}{*}{MNLI}} & \multicolumn{1}{c|}{\multirow{3}{*}{60}} & \multicolumn{1}{c|}{\multirow{3}{*}{50}} & \multicolumn{1}{c|}{\multirow{3}{*}{12.7}} & 1\% & 0.4 & 8.55 & 0.31 & 11.00 & 8.91 \\ \cline{5-10} 
\multicolumn{1}{|c|}{} & \multicolumn{1}{c|}{} & \multicolumn{1}{c|}{} & \multicolumn{1}{c|}{} & 2\% & 0.49 & 8.00 & 0.34 & 10.52 & 8.61 \\ \cline{5-10} 
\multicolumn{1}{|c|}{} & \multicolumn{1}{c|}{} & \multicolumn{1}{c|}{} & \multicolumn{1}{c|}{} & 5\% & 0.65 & 6.89 & 0.47 & 8.37 & 7.34 \\ \hline
\multicolumn{1}{|c|}{\multirow{3}{*}{QQP}} & \multicolumn{1}{c|}{\multirow{3}{*}{60}} & \multicolumn{1}{c|}{\multirow{3}{*}{80}} & \multicolumn{1}{c|}{\multirow{3}{*}{11.3}} & 1\% & 0.25 & 5.84 & 0.12 & 8.88 & 6.41 \\ \cline{5-10} 
\multicolumn{1}{|c|}{} & \multicolumn{1}{c|}{} & \multicolumn{1}{c|}{} & \multicolumn{1}{c|}{} & 2\% & 0.32 & 5.28 & 0.15 & 7.65 & 5.84 \\ \cline{5-10} 
\multicolumn{1}{|c|}{} & \multicolumn{1}{c|}{} & \multicolumn{1}{c|}{} & \multicolumn{1}{c|}{} & 5\% & 0.43 & 4.31 & 0.26 & 5.94 & 4.76 \\ \hline
\multicolumn{1}{|c|}{\multirow{3}{*}{SST-2}} & \multicolumn{1}{c|}{\multirow{3}{*}{60}} & \multicolumn{1}{c|}{\multirow{3}{*}{50}} & \multicolumn{1}{c|}{\multirow{3}{*}{18.4}} & 1\% & 0.23 & 4.30 & 0.09 & 7.78 & 5.25 \\ \cline{5-10} 
\multicolumn{1}{|c|}{} & \multicolumn{1}{c|}{} & \multicolumn{1}{c|}{} & \multicolumn{1}{c|}{} & 2\% & 0.28 & 3.94 & 0.16 & 4.91 & 3.90 \\ \cline{5-10} 
\multicolumn{1}{|c|}{} & \multicolumn{1}{c|}{} & \multicolumn{1}{c|}{} & \multicolumn{1}{c|}{} & 5\% & 0.46 & 2.70 & 0.28 & 3.65 & 3.05 \\ \hline
\multicolumn{1}{|c|}{\multirow{3}{*}{QNLI}} & \multicolumn{1}{c|}{\multirow{3}{*}{60}} & \multicolumn{1}{c|}{\multirow{3}{*}{60}} & \multicolumn{1}{c|}{\multirow{3}{*}{21.5}} & 1\% & 0.18 & 8.46 & 0.13 & 12 & 9.07 \\ \cline{5-10} 
\multicolumn{1}{|c|}{} & \multicolumn{1}{c|}{} & \multicolumn{1}{c|}{} & \multicolumn{1}{c|}{} & 2\% & 0.29 & 7.38 & 0.15 & 10.22 & 8.32 \\ \cline{5-10} 
\multicolumn{1}{|c|}{} & \multicolumn{1}{c|}{} & \multicolumn{1}{c|}{} & \multicolumn{1}{c|}{} & 5\% & 0.44 & 5.89 & 0.25 & 8.01 & 6.85 \\ \hline
\end{tabular}
}
\vspace{5pt}
\end{table}
In order to quantify the different tradeoffs, and evaluate the synergistic impact on the model accuracy from the memory and latency optimizations, the eNVM modeling, and the EE predictor, we implemented the training and evaluation procedures illustrated in Fig.~\ref{fig:train_eval_proc} on the base of HuggingFace’s Transformers infrastructure~\cite{huggingface}.
\subsection{Training and Evaluation Procedure} 
\label{sec:train_proc}

The training methodology consists of two phases. 
In the first phase, the model is pruned during fine-tuning: magnitude pruning is applied to the embedding layer and either movement or magnitude pruning is applied to the Transformer encoder layer. 
An additional loss term comes from knowledge distillation using the base ALBERT model fine-tuned on the target task as a teacher. 
The embeddings and the encoder layer are subject to separate pruning schedules. 
At the same time, the attention heads learn their optimal spans. 
In the second training phase, we freeze the model's parameters prior to fine-tuning the early exit highway off-ramps.

At evaluation time, 8-bit floating-point quantization is applied on all the weights and activations. The quantized embedding weights are modeled according to a 2-bit per cell multi-level (MLC2) ReRAM NVM configuration. The learned attention span mask is element-wise multiplied with the attention weights to re-modulate their saliencies. Entropy prediction is then deployed along with early exit during inference according to Algorithm~\ref{algo:edgebert_hard}.
\subsection{Impact on Model Accuracy, Computation, and Storage} 
\label{sec:benefits}

Using the multi-step procedure illustrated in Fig.~\ref{fig:train_eval_proc}, we amalgamate into ALBERT the various memory and latency reduction techniques at training and evaluation times. Table~\ref{tab:simult_sw} summarizes the generated benefits of the synergistic inference with the following main observations: 
\begin{itemize}
    \item \NAME latency-aware inference provides comparable average exit layer for the same accuracy threshold as the conventional EE approach, while allowing the DVFS algorithm to reduce the frequency and voltage in accordance with the predicted exit layer.
    \item The \NAME approach requires a lower entropy threshold than the conventional EE approach for the same accuracy target; this demonstrates that the we must predict conservatively due to the classification error introduced by the neural network-based entropy predictor. 
    \item Across the four corpora, a uniform 40\% density in the embedding layer is achieved, establishing a compact memory baseline of 1.73MB to be stored in eNVMs.
\end{itemize}


\section{The EDGEBERT Hardware Accelerator System}\label{sec:arch}
\begin{figure}[!t]
    \centering
    \includegraphics[width=\columnwidth]{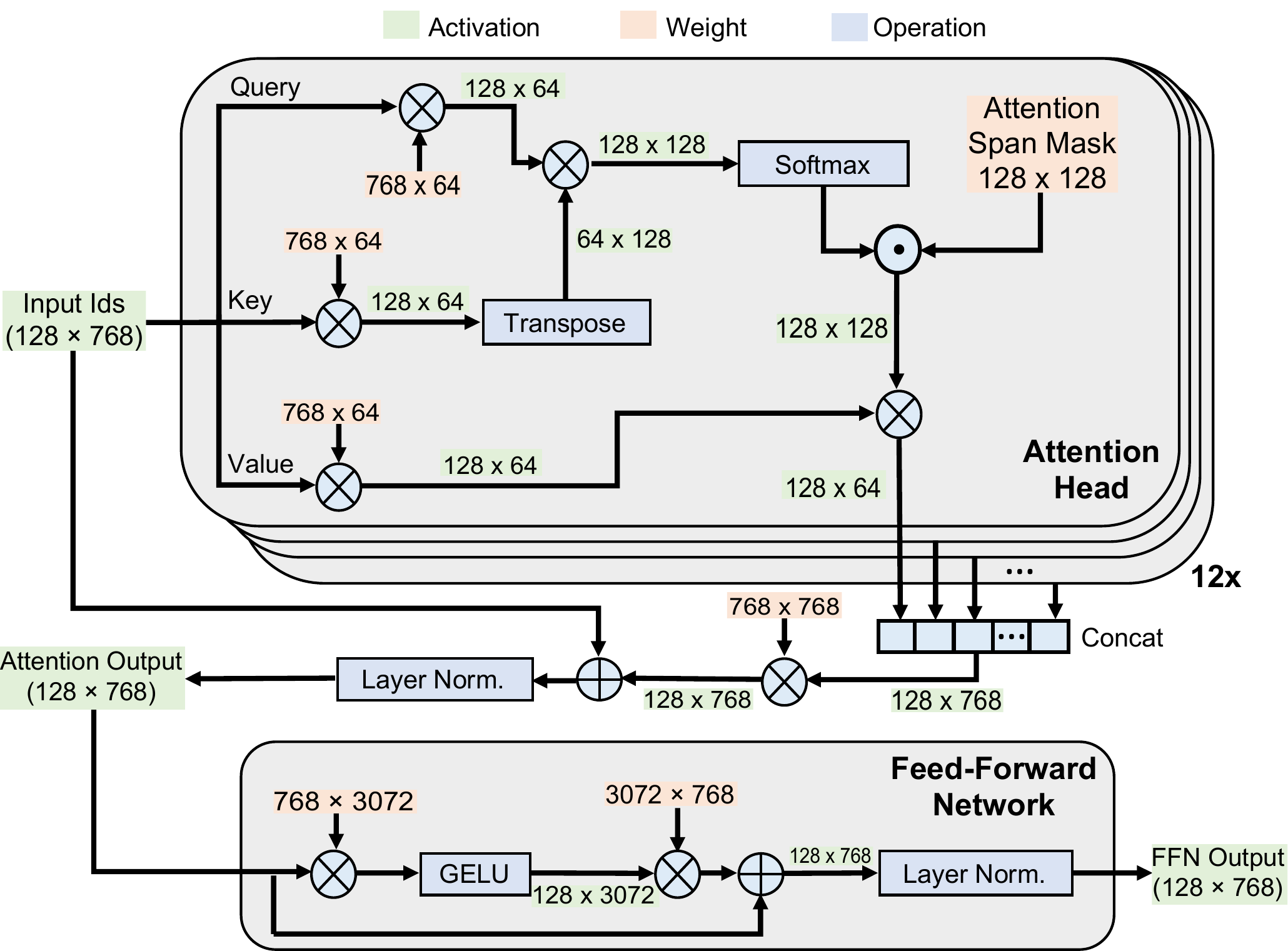}
    \caption{Computations inside the Transformer encoder with attention span modulation. Here, the input sequence is composed of 128 tokens. To simplify the computational diagram, the bias layers are not included.}
    \label{fig:bert_compute}
\end{figure}

\subsection{Required Computations in ALBERT}\label{sec:decompose}
The Transformer encoder is the backbone of ALBERT/BERT, consuming more than 95\% of inference computations. Fig.~\ref{fig:bert_compute} summarizes the computations required in this unit. 
Assuming a sentence length of 128, the transformer encoder requires 1.9GFLOPs to compute matrix multiplications, layer normalizations, element-wise operations (add, mult.), and softmax.
The attention span mask learned during fine-tuning is element-wise multiplied with the softmax output. 
Notably, all the computations inside any of the twelve attention \lpfixme{head} units can be effectively skipped in case its associated attention span mask is 100\% null. The \NAME accelerator reaps this benefit by enforcing adaptive attention span masking during fine-tuning.
\subsection{The \NAME Accelerator System}\label{sec:accel_overview}
\begin{figure*}[t]
    \centering
    \includegraphics[width=0.95\textwidth]{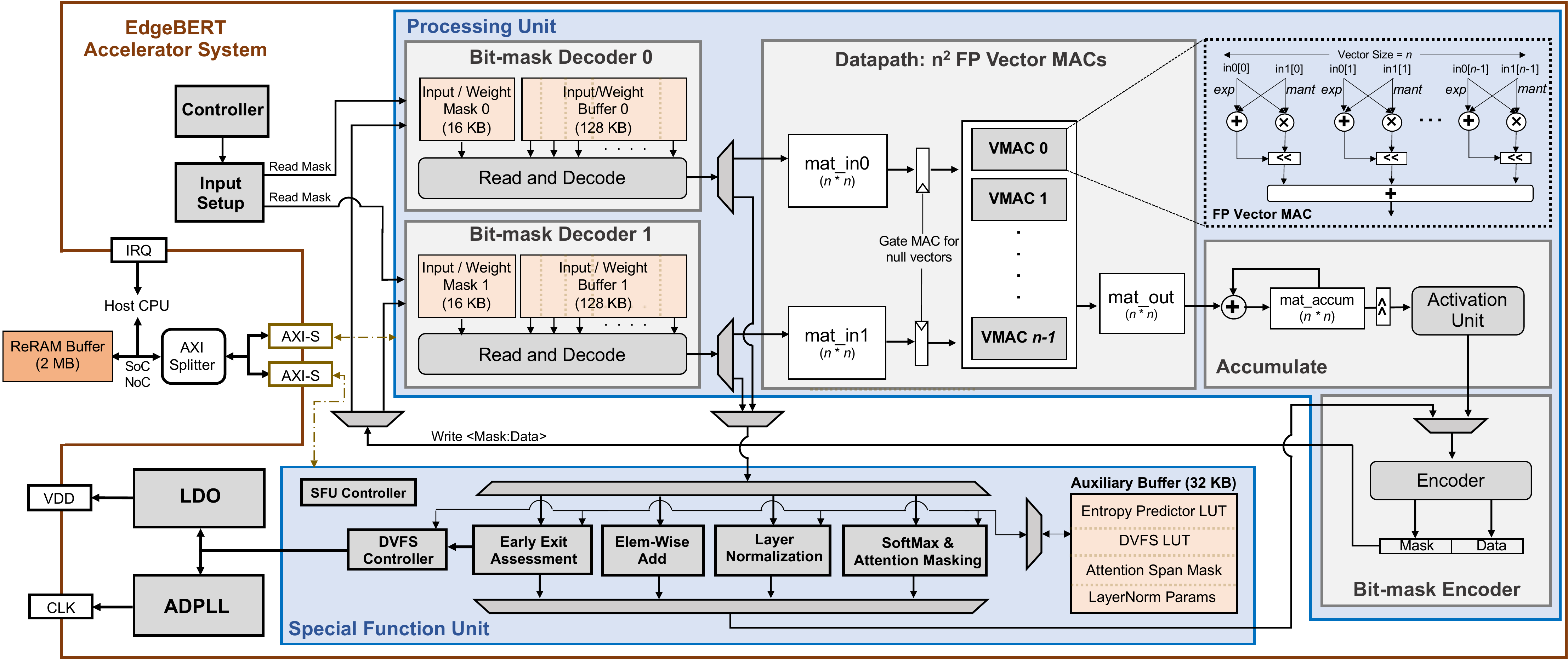}
    \caption{The \NAME hardware accelerator system highlighting its processing unit (PU), and special function unit (SFU). A fast-switching LDO and fast-locking ADPLL are also integrated for latency-driven DVFS.}
    \label{fig:accel}
    \vspace{-1em} 
\end{figure*}

In order to maximize the benefits of the latency and memory reduction techniques during latency-aware inference, we designed a scalable accelerator system that exploits 
these algorithms for compute and energy efficiency \lpfixme{with the following key highlights}: 
\begin{itemize}
\item \lpfixme{Specialized datapath support for \textit{(i)} early exit assessment, \textit{(ii)} softmax and attention span masking, and \textit{(iii)} layer normalization. We notably reformulate their mathematical definitions in order to avoid numerical instability, and where possible, hardware components with long cyclic behaviors such as divisions.}
\item \lpfixme{Non-volatile and high density storage of the shared multi-task parameters substantially improves the accelerator's energy and area efficiency (Sec. \ref{sec:nvm_benefit})}.

\item On-demand DVFS aided by the integration of a fast-locking ADPLL and a fast-switching LDO  regulator.

\item \lpfixme{Compressed sparse execution via bitmask encoding.}
\end{itemize}

The \NAME hardware accelerator, illustrated in Fig.~\ref{fig:accel}, consists of a processing unit (PU), a special function unit (SFU), a LDO and ADPLL for latency-bounded DVFS.  
The communication between the PU and SFU occurs via a custom-built bi-directional streaming channel. An AXI splitter arbitrates the CPU-controlled flow of instructions and data bound for the PU and SFU AXI-slave partitions. 
The multi-task embedding pruned weights and corresponding bitmask are stored in a {2MB} ReRAM NVM buffer in order to avoid reloading them when powered on. Specifically, the bitmask embedding values are stored in a single-level cell (SLC) ReRAM configuration while the nonzero embedding parameters are kept in a 2-bit per cell (MLC2) ReRAM structure, according to the learnings from the NVM studies (Sec.~\ref{sec:nvm}).

\subsection{Processing Unit}\label{sec:PU}

The processing unit (PU) is designed to execute matrix-matrix multiplications in linear layers and attention heads of ALBERT. 

In the PU datapath in Fig.~\ref{fig:accel}, $n$ defines the number of parallel floating-point vector MACs (VMAC) and the vector size of each VMAC. So, there are $n^2$ MAC units in total. The PU datapath takes two $n*n$ matrices as input and computes $n*n*n$ MAC operations in $n$ clock cycles. 
We use 8-bit floating point as the input and weight data type as no accuracy degradation was observed, and 32-bit fixed-point during accumulation. The PU accumulator sums activation matrices and quantizes the final matrix back to 8-bit floating-point. 

To exploit sparsity in both input and weight matrices, we (1) adopt bit-mask encoding and decoding for compressing and decompressing the sparse matrix, and (2) implement skipping logic in the datapath. Bit-masks are binary tags to indicate zero and non-zero entries of a matrix so that only non-zero entries are stored in the decoder SRAM scratchpads. 
For every cycle during decoding, a size $n$ vector is fetched and decoded. The decoder first reads a $n$-bit mask from the single-banked mask buffer to figure out what bank in the $n$-banked input can be neglected, and inserts zero values back to the original zero entries. The encoder also takes a similar approach. It creates a bit mask vector and removes zero entries from the data vector before sending the compressed mask and data vector to one of the PU decoder blocks. 
To save energy, the PU datapath skips the computation of a VMAC product-sum if one of the operand vectors contains only zero values. Although the cycle-behavior of the datapath is not affected by the sparsity of inputs due to the fixed scheduling of data accesses and computations, skipping VMAC operations saves up to {1.65$\times$} in energy consumption (Sec.~\ref{sec:ppa}).

\subsection{Special Function Unit} \label{sec:SFU}
The special function unit (SFU) contains specialized datapaths that compute the EE assessment, DVFS control, element-wise addition, layer normalization, and softmax, all of which get invoked during the latency-aware \NAME inference. The SFU also integrates a 32KB auxiliary buffer to house the EE and DVFS LUTs, the layer normalization parameters, and the multi-head attention span masks learned during the fine-tuning process. \textcolor{black}{All the computations in the SFU are in 16-bit fixed-point format.}
\subsubsection{Computing the Multi-Head Attention} \label{sec:mha}\hfill\ 

\noindent While the linear layers for the attention query, key and value tensors are computed in the PU, the proceeding softmax operation is optimized in the SFU softmax unit.

First, prior to computing an attention head, the SFU controller inspects its associated attention span mask in the auxiliary buffer. 
In case the attention span mask for an attention head is null, the SFU controller proactively cancels and skips entirely the sequence of computations required for that head, and directly writes zero in the corresponding logical memory for its context vector stored in one of the PU decoder blocks.
\setlength{\textfloatsep}{0pt}
\begin{algorithm}[!t]

\small
\scriptsize
\DontPrintSemicolon
\KwIn{attention matrix $A$, and mask $A_M$ of size ($T*T$)}
\KwOut{masked softmax output matrix $A_O$}
$T := \text{number of tokens}$; $n := \text{tile size}$;\


\For{$i = 0$ to $T-1$}{

// Step 1: compute max value\;

    $max = -\infty$\;
    
    \For{$j = 0$ to $T-1$}{
        $vec <= load(A_{[i][n*j:n*j+n-1]})$\;
        
        \If{$max < max(vec)$}{
           $max = max(vec)$\; 
        }
    }

// Step 2: compute log-exponential-sum\;

    $sum_{exp} = 0$\;
    
    \For{$j = 0$ to $T-1$}{
        $vec <= load(A_{[i][n*j:n*j+n-1]})$\;
    
        $sum_{exp} += sum(exp(vec - max))$\;
    }
    
     $logsum_{exp} = ln(sum_{exp})$\;
    
// Step 3: Get softmax and modulate with attn span mask 

    \For{$j = 0$ to $T-1$}{
        $vec<= load(A_{[i][n*j:n*j+n-1]})$\;
        
        $mask <= load(A_{M[i][n*j:n*j+n-1]})$\;
    
        $vec = exp(vec - max - logsum_{exp})$\;
        
        $vec = vec * mask$\;
        
        $store(vec) => A_{O[i][n*j:n*j+n-1]}$\;
    }
}
\caption{{Computing Softmax and Attention Span Masking}}
\label{algo:softmax}
\end{algorithm}

In case the attention span mask for a head contains non-zero elements, the softmax unit takes advantage of the \textit{LogSumExp}~\cite{logsumexp} and \textit{Max}~\cite{max_trick} tricks to vectorize the computation of the softmax function $SM()$ as: 
\begin{equation} \label{eq:softmax}
\resizebox{0.8\hsize}{!}{%
$SM(A_{k})= exp[A_{k} - MAX_{k}(A) 
- ln (\sum^{K}_{k=1}exp(A_{k}-MAX_{k}(A)))]$%
}
\end{equation}
By doing so, the hardware prevents numerical instability stemming from exponential overflow, and avoids the computationally intensive division operation from the original softmax function. 
Upon completing the softmax operation, the softmax unit then performs element-wise multiplication between the resulting attention scores and the attention span mask as described in Algorithm~\ref{algo:softmax}.

\subsubsection{Performing Early Exit Assessment} \label{sec:ee_hw}\hfill\ 

\noindent The EE assessment unit computes the numerically-stable version of the entropy function from equation~\ref{eq:entropy} as follows:
\begin{equation}\label{eq:entropy_hw}
\resizebox{0.8\hsize}{!}{%
$H(x_k) = \ln (\sum\limits_{k=1}^{n} e^{x_k - MAX_{k}(x) }) -  MAX_{k}(x) -
\frac{
  \sum\limits_{k=1}^{n} x_k e^{x_k - MAX_{k}(x)}
}{\sum\limits_{k=1}^{n} e^{x_k - MAX_{k}(x) }}$%
}
\end{equation} 
The EE assessment unit then compares the result with the register value for the entropy threshold. If the EE condition is met, the unit then triggers the accelerator's interrupt (IRQ). Otherwise, the SFU controller initiates the computation of the next Transformer encoder. 
In the case of latency-aware inference in intermittent mode, the EE assessment unit also indexes the EE predictor LUT stored in the auxiliary buffer in order to acquire the predicted exit layer value, which is then passed on to the DVFS controller.

\subsubsection{DVFS System} \label{sec:dvfs_hw}\hfill\ 

\noindent During each sentence inference, the DVFS FSM algorithm keeps track of the EE predictor result and manages the operating voltage and frequency accordingly. Based on the predicted early exit layer, the DVFS controller indexes the logical memory for the $V/F$ LUT table in the auxiliary buffer and extracts the lowest corresponding supply voltage value, $VDD_{opt}$. 
At the same time, the DVFS controller simultaneously updates the ADPLL and LDO configuration registers with settings for $Freq_{opt}$ and $VDD_{opt}$, respectively.
\begin{table}[]
\caption{Performance specs of LDO and ADPLL}
 \vspace{-10pt}
 \label{tab:dvfs_spec}
   \begin{center}
    \begin{small}
     \begin{sc}
 \begin{adjustbox}{width=0.8\columnwidth,center}
\begin{tabular}{@{}cc@{}}
\toprule
LDO response time   & $3.8ns/50mV$ \\
LDO peak current efficiency & $99.2$\% @ $I_{load,max}$ \\
LDO $I_{load,max}$ & $200mA$ \\
ADPLL power & $2.46mW @1GHz$ \\ 
\bottomrule
\end{tabular}
\end{adjustbox}
\end{sc}
\end{small}
\end{center}
\vspace{-5pt}
\end{table}
\begin{figure}[!t]
    \centering
    \includegraphics[width=\columnwidth]{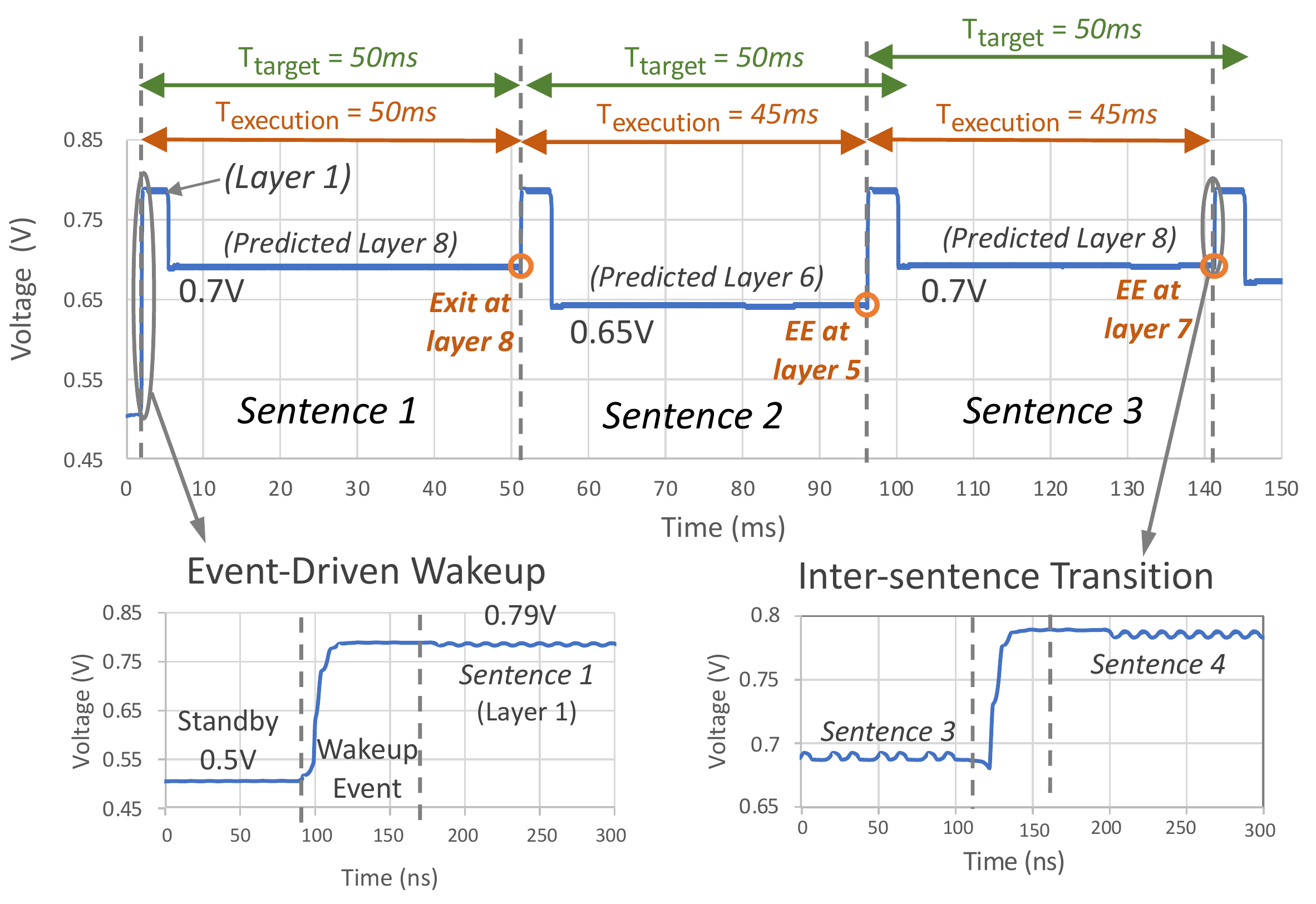}
    \caption{Spice simulations of LDO dynamic voltage adjustments. The LDO stabilizes voltage transitions within 100ns.}
    \label{fig:ldo_sim}
\end{figure}

The synthesizable LDO is implemented using standard power header cells~\cite{2020isscc_intel_ldo}, and evenly distributed across the \NAME accelerator. The LDO is able to scale the accelerator voltage from 0.5V to 0.8V with a 25mV step. With careful power header selection and layout resistance optimization, the LDO can achieve nearly linear scaled power efficiency and a fast response time of 3.8ns/50mV. The ADPLL is also implemented using all-synthesizable approach with the PLL architecture from the FASoC open-source SoC design framework \cite{2020vlsisoc_um_pll}. Following a frequency update request, the all-digital PLL can relock the frequency in a fast speed with low power consumption. The 12nm performance specs of the LDO and ADPLL are shown in Table~\ref{tab:dvfs_spec}.


Fig.~\ref{fig:ldo_sim} show the spice-level simulation of the DVFS for a consecutive sequence of sentence inference. 
For each sentence, the entropy is calculated after the computation of Encoder 1 and sent to the EE predictor to forecast the early exit layer. Based on the predicted early exit encoder and latency requirement for the sentence, the DVFS controller select the lowest voltage level and proper frequency to meet the latency requirement $T_{target}$. Therefore, the remaining encoder stages will compute at a lower voltage level to save energy. For example, the sentence 1 of Fig.~\ref{fig:ldo_sim}, the early exit layer is predicted as 8. Therefore, the rest Encoders (i.e encoder 2-8) in sentence 1 are computed under a lower voltage 0.7V. 

After the inference of the first sentence, the voltage level ramps back to nominal 0.8V for the computation of layer 1 in the following sentence. As on-chip integrated LDO is used, the transition and settling time is optimized to be within 100ns, which is negligible considering the 50ms latency target.
The computation of the next sentence starts once the voltage transition is settled. During idle times, \NAME stays at standby 0.50V to save leakage energy.

\begin{figure*}[!t]
    \centering
    \includegraphics[width=0.98\textwidth]{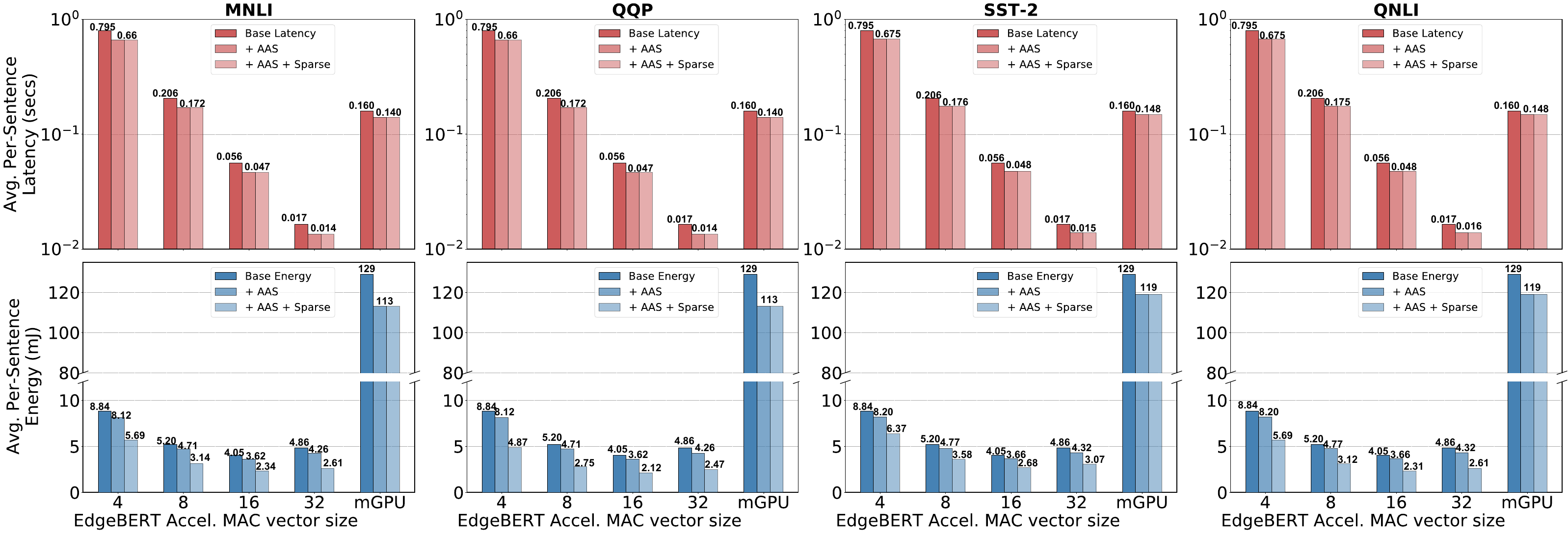}
    \caption{Average latency (top row) and energy (Bottom row) per sentence as the PU MAC vector size scales at max frequency (1GHz) and nominal voltage (0.8V), highlighting impact of adaptive attention span (AAS), and sparsity in weights and activations (Sparse) on the \NAME accelerator and TX2 mGPU. MAC size of 16 yields the most energy efficient design.}
    \label{fig:mac_sweeps}
\end{figure*}
\begin{figure*}[!t]
    \centering
    \includegraphics[width=0.98\textwidth]{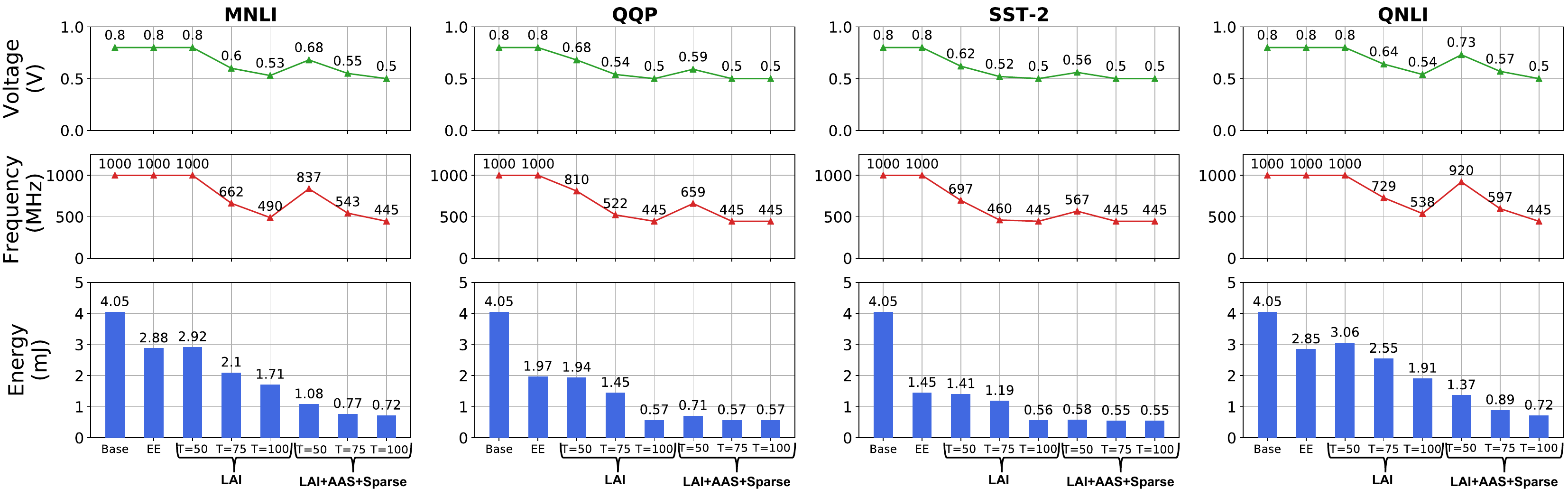}
    \caption{Average DVFS-driven supply voltage (top row) and clock frequency (middle row), as well as, generated energy expenditures (bottom row) of the \NAME accelerator system with $n=16$ during latency-aware inference (LAI), and latency-aware inference further improved with adaptive attention span and sparse execution (LAS+AAS+Sparse). Different latency targets of 50ms (T=50), 75ms (T=75), and 100ms (T=100) are used for LAI executions. Results are compared with the baseline 12-layer inference (Base) and the conventional early exit inference (EE).}
    \label{fig:dvfs_results}
\end{figure*}

\section{Hardware Evaluation}\label{sec:hardware_eval}
\subsection{Design and Verification Methodology}

The \NAME accelerator is designed in synthesizable SystemC with the aid of hardware components from the MatchLib~\cite{matchlib} and HLSLibs~\cite{algoc} open-source libraries. 
Verilog RTL is auto-generated by the Catapult high-level synthesis (HLS) tool~\cite{catapult} using a commercial 12nm process node.
HLS constraints are uniformly set with the goal to achieve maximum throughput on the pipelined design. During the bottom-up HLS phase, the decoder and auxiliary buffers are mapped to synthesized memories from a foundry memory compiler, while the rest of the registers are mapped to D-latches. 
The energy, performance, and area results are reported on the post-HLS Verilog netlists by the Catapult tool at the 0.8V/25c/typical corner.
The 28nm ReRAM cells are characterized in NVSIM~\cite{nvsim} and its read latency, energy, and area are back-annotated into the accelerator results after scaling to a 12nm F$^2$ cell definition in order to match the process node used in the rest of the system.

To quantify the benefits of non-volatility (Sec. ~\ref{sec:nvm_benefit}), we quantify the alternative cost of loading embeddings from off-chip using DRAMsim3~\cite{dram_sim3} to extract cycle-accurate LPDDR4 DRAM energy and latency metrics. GPU results are obtained from CUDA implementations on an Nvidia TX2 mobible GPU (mGPU), whose small form-factor SoC targets embedded edge/IoT applications~\cite{tx2}.


\subsection{Performance, Energy and Area Analyses} \label{sec:ppa}

\subsubsection{Design Space Exploration via MAC scaling} \label{sec:mac_sweep}\hfill\ 

\noindent We start by measuring the energy-performance trade-offs of the \NAME accelerator by scaling its PU MAC vector size. Simultaneously, we further quantify the benefit of bitmask encoding and the predicating logic of the adaptive attention span mechanism by using the attained optimization results (i.e. embedding and encoder sparsity percentage, and attention span) reported in Table~\ref{tab:simult_sw} in which the accuracy drop was at 1\%-pt of the baseline. Adaptive adaptive span is also applied to the mGPU platform in order to quantify and compare the extent of these benefits.

Fig.~\ref{fig:mac_sweeps} shows that the per-sentence processing latency decreases by roughly 3.5$\times$ as the vector size doubles. Across the four tasks, the energy-optimal accelerator design is obtained with a MAC vector size, $n$, of 16. This is because the increase in the datapath power consumption with $n=32$ starts to subdue throughput gains.
The predication/skipping mechanism of adaptive attention span reduces the accelerator processing time and energy consumption by up to 1.2$\times$ and 1.1$\times$, respectively. 
Compressed sparse execution in the PU datapath amounts to an additional 1.4--1.7$\times$ energy savings with QQP receiving the benefit the most.
The \NAME accelerator starts to outperform the mGPU processing time with $n=16$. This energy-optimal design generates up 53$\times$ lower energy compared to the mGPU when all the optimizations are factored in. 

\begin{figure}[!t]
    \centering
    \includegraphics[width=\columnwidth]{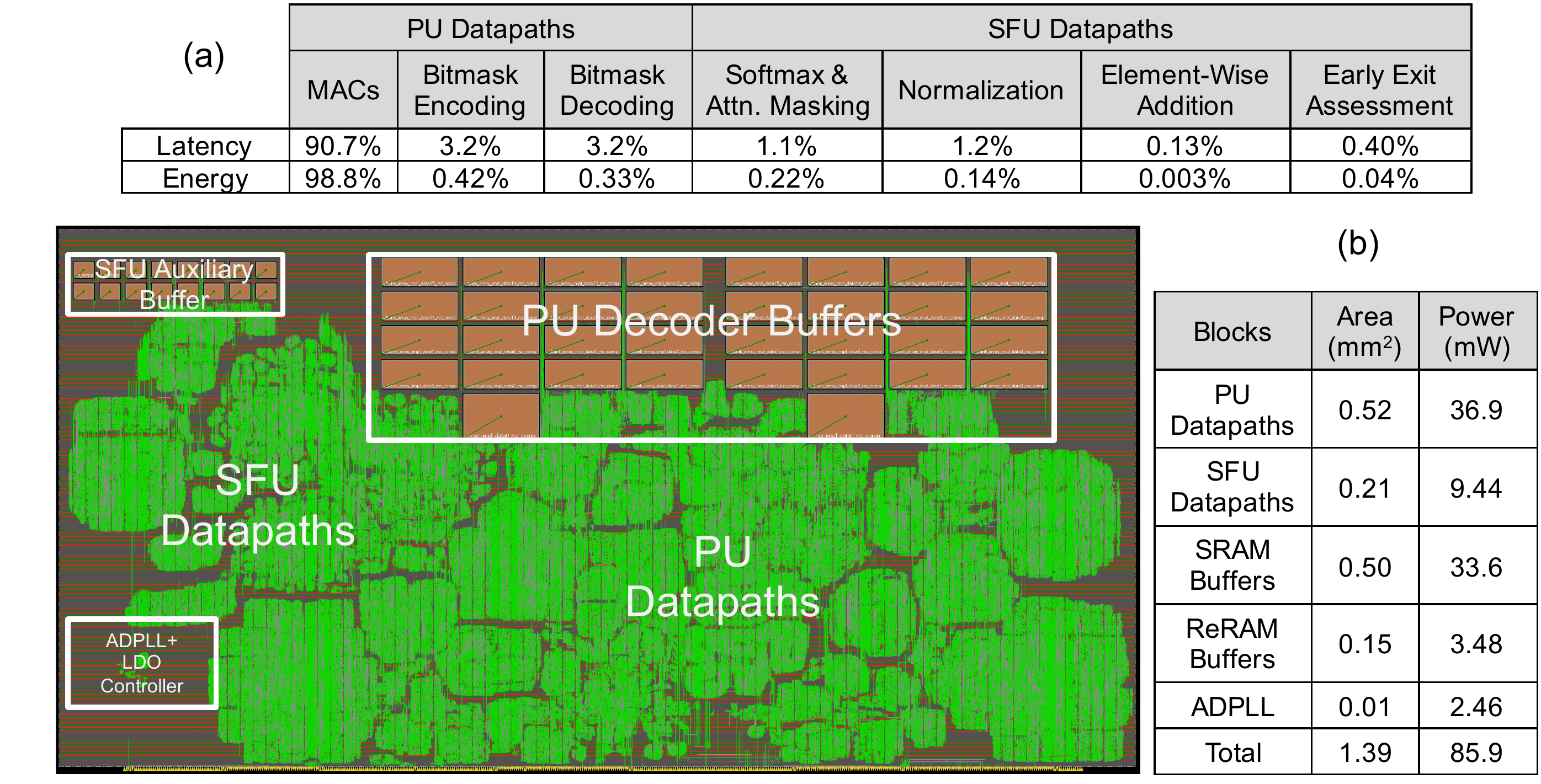}
    \caption{(a) Breakdown of latency and energy consumption in PU and SFU datapaths, and (b) 12nm physical layout, and area and power (@ 0.8V/1GHz) breakdown of the energy-optimal EdgeBERT accelerator (MAC size=16).}
    \label{fig:edgebert_pnr}
    \vspace*{3.5mm} 
\end{figure}

Fig.~\ref{fig:edgebert_pnr} breaks down the latency, energy, area and power contributions inside the placed-and-routed, energy-optimal ($n$=16) \NAME accelerator system which occupies 1.4mm$^2$ while consuming an average power of 86mW. 

\subsubsection{DVFS-based Latency-Aware Inference} \label{sec:lat_hw_results}\hfill\ 

\noindent Fig.~\ref{fig:dvfs_results} shows the DVFS-controlled supply voltage and clock frequency, and the energy savings of the latency-aware inference (LAI) on the energy-optimal accelerator design (i.e. with MAC vector size $n=16$) using latency targets between 50ms and 100ms (common latency thresholds for real-time human perception~\cite{target_latency_blog})). The results show that \NAME optimized LAI achieves up to 7$\times$, and 2.5$\times$ per-inference energy savings compared to the conventional inference (Base), and latency-unbounded early exit (EE) approaches, respectively, as seen in the SST-2 case. As AAS further cuts the number of computation cycles, we observe further relaxation of the supply voltage and clock frequency. At some latency targets (e.g., 75ms and 100ms in QQP and SST-2), further energy savings are not possible as V/F scaling bottoms out. 
\textcolor{black}{To underscore the different contributions to energy savings, at 75ms latency target for example in the case of MNLI, early exit prediction, adaptive attention span, DVFS, sparse execution, and eNVMs account for 21\%, 12\%, 23\%, 39\%, and 5\%, respectively, of the total accelerator energy reduction.}

For stricter latency targets (e.g. $<$ 20ms), the proposed DFVS-based scheme can be used by scaling up to even higher MAC vector sizes (i.e. $n \geq 32$).


\subsection{Benefits of NVM Embeddings Storage} \label{sec:nvm_benefit}

BERT word embeddings are a natural fit for non-volatile storage, given that in \NAME, we freeze them during fine-tuning and reuses them during inference
By virtue of this scheme, we have established a compact 1.73MB baseline wherein the bitmask of the word embeddings is stored in a SLC ReRAM while the nonzero parameters are stored in a 2-bit per cell (MLC2) ReRAM buffer. 

\begin{figure}[!t]
    \centering
    \includegraphics[width=0.95\columnwidth]{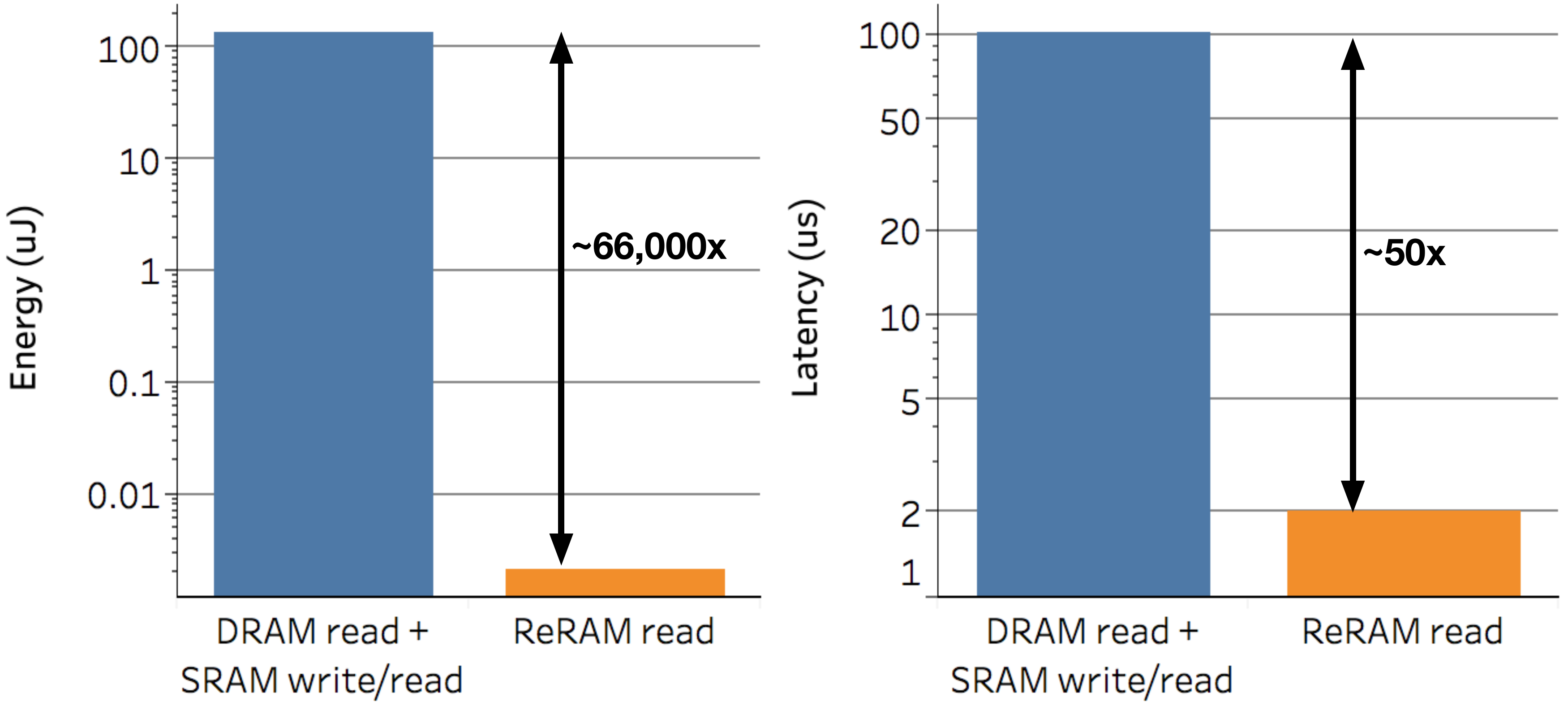}
    \caption{Costs of reading all embedding weights after system power-on. Storing embeddings in ReRAMs gives \NAME significant energy and latency advantages compared to the conventional approach requiring DRAM read followed by SRAM write/read.}
    \label{fig:nvm_benefit}
    \vspace*{3.5mm}
\end{figure}

Fig.~\ref{fig:nvm_benefit} illustrates the immense gains of leveraging this eNVM configuration during single-batch inference after SoC power-on. In \NAME, ALBERT embeddings would only need to be read from the integrated ReRAM buffers due to being statically pre-loaded. The conventional operation dictates reading the embedding weights from off-chip DRAM, then writing them to dedicated on-chip volatile SRAM memories so they can be reused for future token identifications. The \NAME approach enforces a latency and energy advantage that is, respectively, 50$\times$ and 66,000$\times$ greater than the overhead costs in the conventional operation. The non-volatility of this embedded storage means that these benefits can further scale with the frequency of power cycles. 

\section{Related Work}\label{sec:related_hw}

Over the last decade, there has been extensive research on the design of high-performance and energy-efficient DNN hardware accelerators~\cite{eie,minerva,diannao,sm2-jssc,smiv,eyeriss,tabla,fixy2019sysml,snapea,cnvlutin,UCNN,cambricon,gist,Jouppi2017, outlier_aware,scnn,scalesim-ispass,compute_reuse,simba, maeri,pgma-vlsi10,pudiannao,sta_cal20,tangram,smaug-taco,dnnweaver,flexnlp-isscc21}. As these accelerators are increasingly deployed at all computing scales, there is additional interest in the hardware community to automatically generate designs~\cite{magnet,scaledeep,dsagen,chipkit}. However, most of these works focus on CNN and RNN~\cite{tinylstms} computations, and not as much scrutiny has been given to accelerating Transformer-based networks with self-attention mechanisms. 

Recent work in accelerating Transformer-based NLP includes $A^{3}$~\cite{a3}, which proposed a hardware architecture that reduces the number of computations in attention mechanisms via approximate and iterative candidate search. 
\lpfixme{However, the $A^{3}$ scheme fetches the full and uncompressed data from DRAM before dynamically reducing computations in the hardware. In contrast, \NAME learns the optimal attention search radius during the finetuning process and then leverages its very sparse mask to avoid unnecessary matrix multiplications. Therefore, our approach substantially eliminates DRAM accesses as the computation and memory optimizations are pre-learned before hardware acceleration.}

GOBO~\cite{gobo} \lpfixme{focuses on BERT quantization only via} 3-bit clustering on the majority of BERT weights while storing the outlier weights and activations in full FP32 precision. \lpfixme{Although this scheme significantly reduces DRAM accesses, it requires a mixed-precision computational datapath and a non-uniform memory storage. 
In contrast, \NAME adopts uniform 8-bit data storage in SRAM and eNVMs memories.}   
Lu~{\it et al.}\ \cite{MHA_accel} proposes a dense systolic array accelerator for the Transformer's multi-head attention and feed-forward layers and optimizes Transformers' computations via matrix partitioning schemes. The \NAME accelerator executes compressed sparse inference for higher energy efficiency.
OPTIMUS~\cite{optimus} looks to holistically accelerate Transformers with compressed sparse matrix multiplications and by skipping redundant decoding computations. 
FlexASR~\cite{flexnlp-isscc21} accelerates attention-based RNNs in a specialized attention datapath and only saves energy by gating the MAC when decoder RNN inputs are null.
SpAtten~\cite{spatten} accelerates Transformer-based models via progressive cascade token and attention head pruning. The importance of each attention head is determined during the computation via a top-k ranking system. In contrast, \NAME opts to learn the important attention heads during the fine-tuning process by activating adaptive attention spanning. The optimized and sparse attention spans are then used by the \NAME accelerator to predicate the NLP computation.

\begin{figure}[!t]
    \centering
    \includegraphics[width=\columnwidth]{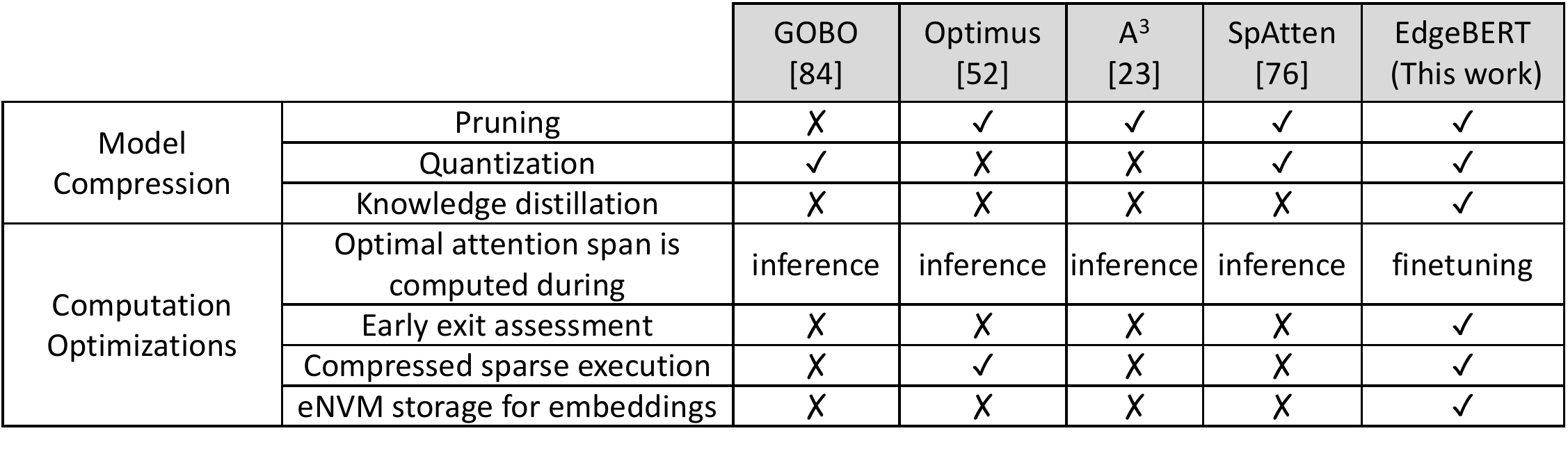}
    \caption{\textcolor{black}{Comparison of \NAME with prior work accelerating Transformer-based NLP models.}}
    \label{fig:prior_work}
\end{figure}

\lpfixme{Finally, all the aforementioned NLP accelerators stores the embedding weights in traditional volatile SRAM memories. By contrast, this work recognizes that embedding weights do not change across NLP tasks. Therefore, \NAME statically stores the word embeddings in high density eNVMs, generating substantial energy and latency benefits (Sec.~\ref{sec:nvm_benefit}).}
Fig.~\ref{fig:prior_work} qualitatively contrasts some of the prior work with \NAME.
\section{Conclusion}\label{sec:conclusion}
As newer Transformer-based pre-trained models continue to generate impressive breakthroughs in language modeling, they characteristically exhibit complexities that levy hefty latency, memory, and energy taxes on resource-constrained edge platforms. 
\NAME provides an in-depth and principled latency-driven methodology to alleviate these computational challenges in both the algorithm and hardware architecture layers. 
\NAME adopts first-layer early exit prediction in order to perform dynamic voltage-frequency scaling (DVFS), at a sentence granularity, for minimal energy consumption while adhering to a prescribed target latency. 
Latency and memory footprint overheads are further alleviated by employing a balanced combination of adaptive attention span, selective network pruning, floating-point quantization.
We further exploit and optimize the structure of eNVMs in order to store the shared multi-task parameters, granting \NAME significant performance and energy savings from system power-on. 
Sentence-level, latency-aware inference on the \NAME accelerator notably consumes 7$\times$ and 2.5$\times$ lower energy than the conventional full-model inference, and the latency-unbounded early exit approach, respectively.

\section*{Acknowledgement}
This work was supported in part by the Center for Applications Driving Architectures (ADA), one of six centers of JUMP, a Semiconductor Research Corporation (SRC) program co-sponsored by DARPA; DARPA’s DSSoC program; NSF Awards 1704834 and 1718160; Intel Corp.; and Arm Inc.



\bibliographystyle{IEEEtranS}
\bibliography{refs}

\end{document}